\documentstyle[12pt]{ioplppt}

\begin{document}

\jl{3}

\newcommand{\taisy}{(TaSe$_4$)$_2$I}
\newcommand{\cdw}{charge density wave}
\newcommand{\lra}{Lee, Rice and Anderson}
\newcommand{\qod}{quasi one-dimensional}
\newcommand{\xps}{core hole photoemission}
\newcommand{\tc}{\mbox{263 K}}
\newcommand{\tdot}{T_{\mbox{\tiny 3D}}}
\newcommand{\tcmf}{T_{\mbox{\tiny MF}}}
\newcommand{\kB}{k_{\mbox{\tiny B}}}
\newcommand{\taulra}{\tau_{\mbox{\tiny LRA}}}
\newcommand{\psisq}{\langle \psi^2 (T) \rangle}
\newcommand{\invxi}{\xi^{-1}(T)}
\newcommand{\invxisq}{\xi^{-2}(T)}
\newcommand{\doslra}{D_{\mbox{\tiny LRA}}}
\newcommand{\iomegan}{\i \omega_{\rm n}}
\newcommand{\ie}{i.\ e.\ }
\newcommand{\eg}{e.\ g.\ }
\newcommand{\be}{\begin{eqnarray}}
\newcommand{\en}{\end{eqnarray}}
\newcommand{\no}{\nonumber}
\newcommand{\hc}{{\it h . c.}}

\title{Electronic properties of the
pseudogap system \taisy}[Pseudogap in (TaSe$_4$)$_2$I.]

\author{Nic Shannon and Robert Joynt}

\address{Department of Physics, University of Wisconsin-Madison,
1150 University Avenue, Madison, WI 53706, USA}

\begin{abstract}

The room temperature ``metallic''  properties of the
quasi-one-dimensional material \taisy\ differ markedly from those
expected of either a Fermi or a Luttinger Liquid, showing strong signs of
a suppression of the density of states at the Fermi level.  We present 
evidence for the existence of strong quasi--static fluctuations
of structural order with long correlation length.  
These fluctuations produce a pseudogap in the density of 
states.  We compute the temperature dependence of the
optical and DC conductivities of \taisy\ in its conducting phase, the
nature of its core hole spectra, the NMR Knight Shift and relaxation rate.
Predictions for these quantities are made on the
basis of a \lra\ model.  This model represents the simplest theory
of a pseudogap, and gives satisfactory agreement with experiment in 
the cases where comparisons can be made.  In contrast, the predictions of a
strongly correlated (Luttinger Liquid) model appear to to contradict
the data.  The chief remaining discrepancy is that the gap deduced from
transport quantities is less than that observed in 
photoemission.  We discuss some possibilities for resolving this issue.

\end{abstract}


\maketitle

\section{Introduction}

\taisy\ is among the most widely studied of \qod\ materials.  First
synthesized in 1984 \cite{gressier1}, it provides a good example of a
``metallic'' system undergoing a Peierls transition to a nearly
commensurate \cdw\ ground state.  It is unusual among low dimensional
systems in the
uncommonly high temperature (\tc) at which it becomes ordered, and in
the way in which that ordering affects its room temperature
properties.  It has attracted interest as a Luttinger liquid candidate,
based on its photoemission spectrum \cite{dardel}, \cite{terrasi}.  
We argued in an earlier paper \cite{shannon}
that none of the observations on \taisy\ are really consistent with the 
Luttinger picture, and that the photoemission is best explained
by its proximity to a \cdw\ state --- the energy
scale associated with the gap which opens in the \cdw\ state is
clearly visible in the room temperature properties of \taisy, 
as was shown by early
measurements of magnetic susceptibility \cite{johnston1}.
{\it \taisy\ is clearly a 'pseudogap' system.}

Below \tc, the resistivity of \taisy\ shows an activated behavior, and
measurements of its optical conductivity and of the
dispersing features measured in angle-resolved photoemission
(ARPES) spectra also give evidence of a
gap. Strangely however, the gap (or, above \tc\, the pseudogap)
measured by different experiments is {\it not}
of the same size, as is illustrated in Table 1, where gaps
and their corresponding
mean-field transition transition temperatures are listed.

\begin{table}
\caption{Estimates of gap size and transition temperatures.}
\lineup
\begin{indented}
\item[]\begin{tabular}{llll}
\br
 &$\Delta$&$\tdot$&$\tcmf$\\
\mr
DC Conductivity (e.g. \cite{terrasi})&180 meV/2100 K&263 K&---\\
Optical Conductivity \cite{berner}&200 meV/2300 K&---&---\\
ARPES \cite{shannon,terrasi} &520 meV/6000 K&---&892 K\\
Magnetic Susceptibility \cite{johnston1}&---&---&860 K\\
\br
\end{tabular}
\end{indented}
\end{table}

Roughly speaking, the gap in ARPES spectra is 2 to 3 times that
in other experiments.
We discuss below how this discrepancy may be resolved.

Coincidentally, observation of the pseudogap state in underdoped high-T$_c$
superconductors has caused great excitement.  The nature of this
state is presently intensely debated, with no consensus
having been reached.  In this context, it is important to look in
some detail at \taisy\ which, we believe, is a much
simpler pseudogap material.  In fact we will contend that
its properties, with the single exception of the gap discrepancy, 
can be understood in a straightforward, and even rather crude, theory.

The dispersing features measured by ARPES in the
conducting phase were found to be quite well described by a 
simple \lra\ (LRA) model in which
the fluctuations of lattice order associated with the Peierls
transition are viewed
as a temperature dependant ensemble of static potentials in which the
electrons move \cite{lra}, and electron--electron interaction is explicitly 
not included.

In this paper we extend the analysis begun in \cite{shannon}
to optical and DC conductivities, making predictions for the
temperature dependence of these quantities in the conducting phase
on the basis of the same simple LRA model.
We will also discuss what might be learned from NMR and \xps\
experiments on \taisy, considering in particular what might be 
established about spin--charge separation from the comparison of the two.   
In all cases we compare these predictions both with existing experimental 
data and with the behavior expected of models based on strong 
electron--electron interaction.  For greater
readability, ancillary technical details are given in appendices. 

\section{The \lra\ model\label{model}}

Conduction in \taisy\ takes place along one-dimensional
chains of Ta atoms \ref{fig:taisy}.  The material undergoes a
Peierls transition at $T_c = 263 K$ to a state 
where the lattice is distorted by
a condensed transverse acoustic phonon mode.
This may be thought of in simple terms as a
slightly incommensurate tetramerization of the tantalum atoms
perpendicular to the chain axis.  Superlattice
reflections appear in X--Ray spectra at 
$\vec{q} = (\pm 0.05, \pm 0.05, \pm 1.085)$,
confirming incommensurate \cdw\ order not quite aligned with the
chain axis
\cite{qcdw}.  In the chain, this corresponds very nearly to a $2k_F$
fluctuation.
 
The proper model for what is then a one-dimensional 
electron--phonon problem is given by the
Fr\"{o}hlich Hamiltonian
\begin{eqnarray}
H &= & \sum_{k} \epsilon (k) c^{\dagger}_k c_k
     + \sum_{q} \omega (q) b^{\dagger}_q b^{}_q
     + \frac{1}{\sqrt{L}} \sum_{q,k} g(q) c^{\dagger}_{k+q} c^{}_k
u_q,
\label{eqn:froh}
\end{eqnarray}
where
\begin{eqnarray}
u(q) &= & \frac{1}{\sqrt{2\omega (q)}} ( b^{\dagger}_q +b_{-q} ),
\nonumber
\end{eqnarray}

\noindent
and $c^{\dagger}_{k}$ and $b^{\dagger}_{q}$ are (respectively)
creation operators for electrons
and phonons with dispersion $\epsilon (k)$ and $\omega (q)$. $u_q$ is
the Fourier transform of
the lattice displacement, and $g(q)$ the electron-ion coupling.
Since electron spin enters into the problem only in appropriate
factors of two, it will be suppressed in our notation.

The physics of this Hamiltonian
has been widely studied for many years and is
highly nontrivial.  It is well known both that the one dimensional
lattice is unstable against distortion, and that thermodynamic
fluctuations
prevent a transition to a state with long range order from occurring
at finite
temperature in any {\it truly} one--dimensional system.

The possibility of phonons with wave number $Q=2k_F$ decaying into
zero energy particle--hole pairs leads to softening of the phonon
spectrum for $Q\approx 2k_F$, and at a mean field level the system
described by the one--dimensional Fr\"{o}hlich Hamiltonian Eq.\ \eref{eqn:froh}
undergoes a transition to a fully ordered \cdw\ state at the temperature
$\tcmf$ for which the frequency of the $Q=2k_F$ phonon mode goes to zero
\cite{rice}.
This temperature is determined by the strength of electron--phonon
coupling $g(2k_F)$, and is of the same order as the gap developed in the \cdw\
state at absolute zero.  Within the set of approximations usual for BCS theory
one has $2 \Delta = 3.5 \kB \tcmf$ but for real materials this
relation is seldom exact.   The three-dimensional \cdw\ transition
which occurs in \taisy\ and other similar compounds is stabilized 
by interaction between different metallic chains, but occurs at 
a temperature much smaller than the gap energy,
as can be seen in Table 1 where $\tcmf$ is certainly well above 
the exerimental transition temperatue.

In the spirit of LRA, we identified the transition temperature
$T_c = \tdot = 263K$ with a crossover from a three dimensional 
Peierls--distorted mean field \cdw\ state described by
\begin{eqnarray}
\label{eqn:bcsH}
H_{MF} &=& \sum_{k} \epsilon (k) c^{\dagger}_k c_k
     + \sum_{k} [\Delta^* c^{\dagger}_{k-2k_F} c_k
        + \Delta c^{\dagger}_{-k+2k_F} c_{-k}]\\
\Delta &=& \frac{1}{\sqrt{L}} g(2_{k_F}) \langle u_{2k_F}\rangle .
\end{eqnarray}

\noindent
to a state in which there are essentially 
uncorrelated fluctuations of charge density wave order on individual chains. 
We make no attempt to accurately describe the way in which
this dimensional crossover takes place.
Fluctuation effects above $\tdot$ are taken into account in the
simplest possible way consistent with a mean--field ground
state, replacing the order parameter for the Peierls distortion
of the lattice 
$\Delta = \frac{1}{\sqrt{L}} g(2_{k_F}) \langle u_{2k_F}\rangle$
by a static external field which has a non--vanishing expectation
value for other momenta not exactly equal to $2k_F$:
\begin{eqnarray}
\label{eq:lraH}
H_{LRA} &=& \sum_{k} \epsilon (k) c^{\dagger}_k c_k
  +  \sum_{Q, k'>0}
   [ \Psi^*_{Q} c^{\dagger}_{k'-Q} c^{}_{k'}
   + \Psi_{Q} c^{\dagger}_{-k'+Q} c^{}_{-k'} ]  \\
   \Psi_Q &= & \frac{1}{\sqrt{L}} g(Q) \langle u(Q) \rangle.
\end{eqnarray}
The classical field $\Psi_{Q}$ belongs to a thermal ensemble of 
potentials characterized by a mean sqaure ``gap'' scale $\psisq$
reflecting the size of fluctuations of lattice disorder, and 
an inverse coherence length $\invxi$.

Since the main purpose of this paper is to explore the LRA
model as a phenomenology for \taisy\ we consign the derivation 
of the single-particle Green's function for electrons moving in this static 
phonon field to \ref{whatislra}, and reproduce here only the result
\begin{eqnarray}
{\cal G}(k, i\omega_n)^{-1} &=& i\omega_n - \epsilon(k)  -\frac{\psisq}
           {\iomegan + \epsilon (k) \pm \i v_f \invxi} ,
\label{eqn:selfeng}
\end{eqnarray}
where the choice of sign $\pm \i v_f \invxi$ is made according to
whether we continue to the upper or lower half plane.
For convenience we have set $\hbar = 1$.   

Determining the temperature dependence of  $\psisq$ and $\invxi$ forms an 
essentially independent problem; for a fully self-consistent phemonemology 
they should be found from experiment.  The scale of  
$\psisq$ is set by the size of the mean gap at $T = 0$, and it varies 
little over the range of temperatures in which we can apply the LRA 
theory to experiments on \taisy, \ie $263 < T < 430K$.  In contrast $\invxi$
varies quite strongly with temperature, and becomes extremely small 
at the transition temperature $\tdot$.   For all experimentally
accesable temperatures $v_f\invxi << \sqrt{\psisq}$.  Our parameterization 
of the model is discussed in more detail in \ref{param}.  

In what follows we will use a spectral representation of the electrons
\begin{eqnarray}
A (k, \omega) =  \frac{2 v_F \invxi \psisq}
  {[\omega^2 - \epsilon(k)^2  - \psisq]^2 + v_F^2 \invxisq
   [\omega - \epsilon(k)]^2},
\label{eqn:A}
\end{eqnarray}

\noindent
All the anomalous features of the room temperature ARPES data
are present in this spectral function.  In particular
it predicts the broad dispersing features, and ``quasigap'' structure
observed by Terassi \etal \cite{terrasi}.

It is important to note that the results of this paper assume the
validity of \eref{eqn:A} and nothing further, since all experimental 
quantities can be expressed as integrals over the spectral functions.  
Since this equation can be directly compared with experiment, we may 
regard our theory in two different ways.  On the one hand, it may be 
viewed as a phenomenology in which the spectral function is taken from
one experiment (ARPES) and used to develop a picture of several experiments, or 
it may be viewed as a direct test of the LRA model against
experiment, without prejudice as to the suitability of LRA as a solution 
to the Fr\"ohlich Hamiltonian.

\section{Previous Work}

For completeness, we review previous work
on the application of these equations to explain 
experimental data.  An example of a fit to a room temperature ARPES spectrum
for \taisy\ made on this basis in Ref.\ \cite{shannon} is shown in
\Fref{fig:edc}.   A fit to the temperature dependence 
of the uniform magnetic susceptibility of \taisy\ by 
Johnston et al. \cite{johnston1}, is given in \Fref{fig:chi}.
In both cases, the agreement of theory and experiment is excellent.

\section{DC Resistivity}

The DC resistivity $\rho(T)$ of \taisy\ is usually presented in an
Arrhenius plot so as to extract an ordering temperature $\tdot$ and
an activation energy $\Delta_0$ for the \cdw\  state.  For
temperatures above $\tdot$ the data plotted in this 
form show {\it no upturn} ---
in fact the resistivity becomes nearly temperature independent, decreasing
slowly over all higher measured temperatures \cite{johnston2}.   
This contradicts our expectations of an ordinary metal, 
where most scattering mechanisms increase in
effectiveness with increasing temperature, and an upturn in $\rho(T)$
is expected on the closing of the gap.

In the LRA liquid we anticipate that scattering of
electrons from fluctuations of charge density order will have two
effects on conductivity: the suppression of the density of states at
the Fermi energy, and the imposition of a finite lifetime on electrons
propagating in momentum eigenstates along the chains.  
The former is reflected in the pseudogap visible in 
photoemission spectra and the spectral function
\Eref{eqn:A}, the latter in the large temperature-dependent width of
the dispersing features in the spectral functions.
The temperature dependence of the conductivity of
will therefore depend on the interplay between
the gap and lifetime effects as parametrized by $\psisq$
and $\invxi$.

We calculate the intrinsic DC conductivity of an
LRA liquid directly from a current--current correlation function.
The results of the usual Kubo formalism, evaluated to first order,
may be expressed directly in terms of the spectral function of the
system under consideration as \cite{mahan-transport} :
\begin{eqnarray}
\sigma_0^{(1)} = - \frac{\e^2}{2 m^2}
   \int^{\infty}_{-\infty} \frac{\d \omega}{2 \pi}
   \frac{\partial n_f(\omega)}{\partial \omega}
   \sum_k k^2 [A(k,\omega)]^2
\label{eqn:kubo}
\end{eqnarray}
We evaluate the sum over $k$ as a contour integral and perform the
final integral over $\omega$ numerically.   
We do not include disorder in 
our calculations in this paper, and the DC conductivity
is of course sensitive to disorder.  We therefore add a
small temperature independent part to the 
inverse coherence length $\invxi$,
which we chose to be $\xi^{-1}(300 K)$.
This affects the temperature dependence of
the conductivity only near $\tdot$.
In order to present consistent results we
will include this correction in later calculations of optical
conductivities.

Theory and experiment are shown in the interesting 
range from $264 K$ to $430 K$ in Fig.\ \ref{fig:sigmazero}.
There is qualitative agreement, in that
the unusual slow decrease with temperature is reproduced.
However, it is clear that a proper treatment of the crossover
to three dimensions is needed to fit the data near $T_{3D}$.

To understand 
the qualitatively behavior of $\rho (T)$,
we can make an estimate of the scale of the
temperature dependence 
somewhat in the spirit of the Drude model :
\begin{eqnarray}
\sigma(T) \sim \doslra (\epsilon_f) \times \taulra ,
\label{drude1}
\end{eqnarray}

\noindent
where $\taulra$ is the effective lifetime of the electrons
and $\doslra (\epsilon_f)$, the density of states at the Fermi energy,
has been chosen as a representative measure of the degree to which the
gap has filled.

An expression for $\doslra (\omega)$ can be found analytically from
\eref{eqn:A}; for $\omega = \epsilon_f$ it has the simple form
\begin{eqnarray}
\doslra(\epsilon_f) = \frac{v_f \invxi}
       {\psisq + v_f^2 \invxisq} \approx \frac{v_f \invxi}{\psisq}
\label{eqn:dosef}
\end{eqnarray}

\noindent
Naively, we might expect $1/\taulra$ to be given by the imaginary
part of
the self energy of an electron at $k=k_f$, $\omega=0$, but this
diverges for
$\xi \rightarrow 0$, while the reciprocal of the real electron
lifetime
should tend to zero.   However in the limit where $v_f
\invxi/\sqrt{\psisq}
\rightarrow 0$
it is possible to show that the spectral function of Eq.\
\eref{eqn:A}) can be
rewritten
as the sum of two Lorentzians, with $v_f \invxi/2$ playing the role of
a lifetime (peak width).  A naive prediction for the DC conductivity
would then be :
\begin{eqnarray}
\sigma(T) \sim \frac{1}{\psisq}.
\label{drude2}
\end{eqnarray}

\noindent
Since $\psisq$ decreases only slightly over experimentally relevant
temperatures, this represents a weakly temperature-dependent
increasing conductivity, as is observed.
But note that the
conductivity we calculate is anything but temperature--independent over the remainder of the
(greater) range of temperatures which we may access in our model.
By fitting the numerical results we find it is rather well described
by a high
power law in the reduced temperature $(T/\tcmf)^x$, $x\approx 10$, with the
weak temperature dependence of the conductivity at low temperatures being
due to its suppression by many factors of $T/\tcmf$.

Thus the 
very weakly temperature dependent DC conductivity is well explained by a 
cancellation of density-of-states and lifetime effects.  This 
fails to capture the
interesting step-like feature seen at the transition
temperature in an Arrhenius plot.  This is not surprising, as the 
dimensional crossover is not included in a realistic fashion in our 
theory.  
This (near) independence of DC conductivity on temperature
over so large a range of temperatures is sufficiently unusual that we
consider it to be a striking confirmation of the LRA theory.

\section{Optical conductivity of the LRA Liquid}

The optical conductivity of \taisy\ has been measured over a range of
temperatures \cite{berner} and also shows a surprising lack of 
evolution with temperature.

A truly gapped state, such as that described by Eq.\ \eref{eqn:bcsH}
cannot support charge carrying excitations (or indeed any excitations)
at energies less than the $2\Delta$ needed to promote charge across
the gap.   The measured optical conductivity should therefore ``turn
on''
abruptly at $\omega \approx 2\Delta$.  Since the density of states
diverges immediately above and below the gap, we expect a very
pronounced asymmetric peak at $\omega = 2\Delta$ with almost all
weight on the high energy side.

Such a peak is indeed observed in the \cdw\ state of \taisy.  
However, it is rather less asymmetric than might have been expected.
Away from the maximum the spectra grow as $\omega^2$ on the low-- and
fall away as $1/\omega^3$ on the high--energy side.
Since we are only interested in the excitations from the band of
electrons which carry currents in \taisy, we will ignore the
presumably interband contribution which sets in beyond $1.2eV$.

In a mean-field-like picture, we would expect the 
feature at $\omega \approx 2\Delta \approx 400meV$ to
become weaker, broader, and migrate to lower energies 
as the temperature is raised and
the gap closes.  The first two of these expectations are fulfilled,
but over the range of temperatures $15K-400K$ the position of the
maximum changes by a few $meV$ only, and not by the $\approx 120meV$
that might have been expected from our parameterization of the model.

The only qualitative change in the peak as \taisy\ is heated through
$\tdot$ is the loss of the simple $\omega^2$ behavior on the low
energy side --- on a log--log scale the peak grows a small shoulder
at a litlle below $\Delta$.  On the high energy side the peak still
falls away as $1/\omega^3$.

In fact the Hamiltonian in Eq.\ \ref{eqn:bcsH} is not adequate by
itself to
describe the dynamics of the ordered state.  The lattice distortion
may vary slowly in space and time \cite{gruner}, leading to low energy
collective modes in the \cdw\ which are observed at finite frequency
in Blue Bronze {\it even in its conducting
phase} \cite{gorshunov} .  Since we are mostly concerned
with pseudogap effects in the single--electron properties of \taisy\
however,
we will not consider such collective excitations here.

Under this approximation, the optical conductivity of the \lra\ model
can be calculated directly
from a Kubo formula expressed in terms of electron Green's functions,
in the same way as the DC conductivity.  In this
case we start from
\begin{eqnarray}
\sigma(\omega) = - \frac{\e^2}{2 m^2}
   \int^{\infty}_{-\infty} \frac{\d \epsilon}{2 \pi}
   \left[ \frac{n_f(\epsilon+\omega) - n_f(\epsilon)}{\omega} \right]
   \sum_k k^2 A(k,\epsilon) A(k,\epsilon + \omega)
\label{eqn:kubo2}
\end{eqnarray}

\noindent
which obviously has Eq.\ \eref{eqn:kubo} has its $\omega \to 0$ limit.
Once again we perform the sum over $k$ analytically and
the integration over $\omega$ numerically.  Typical
results are presented on a log--log scale in \Fref{fig:optical}.  They
may be characterized by

\begin{itemize}

\item Lorentzian peak at $\omega = 0$ with width of the order
      the naive quasiparticle inverse--lifetime $v_F \invxi$.

\item Strongly suppressed response over the range of energies
      $\omega \approx v_F \invxi \to \sqrt{\psisq - v_F^2 \invxi^2}$.

\item Shoulder at $\omega \approx \sqrt{\psisq - v_F^2 \invxi^2}$ (for
      temperatures sufficiently close to $\tdot$).

\item Sharp asymmetric peak at $\omega \approx \sqrt{\psisq - v_F^2
\invxi^2}$
      with temperature dependent position, width and height.

\item Asymptotic decay as $1/\omega^3$ at high frequencies.

\end{itemize}

These features of the spectra and their temperature dependences
are again in good qualitative agreement with those taken
on \taisy, although it is important to note that
we have been forced to abandon the value of the
zero temperature gap $\Delta_0 = 0.52 eV$ required to fit
photoemission in favour of the smaller value $\Delta_0 = 0.2 eV$
suggested by transport measurements.  This value is listed in Table 1.
We have explicitly checked that the $\omega \to 0$ limit of the
predicted optical conducivity matches our prediction for the 
DC conductivity, providing a useful self--consistency check 
on each calculation.

The important differences been model and experimental spectra are

\begin{itemize}

\item Greater temperature dependence of the peak position.

\item Lack of weight within the gap --- experimental spectra have
      more weight within the ``gap'' especially at low temperatures.

\end{itemize}

Interestingly recent higher resolution photoemission spectra for
\taisy\ also show less temperature dependance in the peak position 
than the earlier data which we fitted in terms of the LRA model
\cite{marco}.  Without attempting a detailed analysis, we note that optical
conductivity  spectra taken on Blue Bronze are qualitatively very similar to those
taken on \taisy.  Spectra for the family of organic \qod\ systems
$(TMTSF)_2 X$, on the other hand, show important differences in detail
\cite{tmtsf}.

We wish to stress that measurement of the different gap scales
by different experiments, and the excess of weight in the gap are 
problems not of the conducting phase only, but also need to be addressed 
in the ordered phase.
To do this it may be necessary a) to go beyond the approximation 
of static lattice distortion which underlies both the mean field theory
of the ordered state and the LRA model and b) to include the effects 
of electron--electron interaction and/or impurities in a more careful way.   

One way of obtaining a better fit to optical conductivty spectra in 
the ordered phase would be to convolute the expected mean field 
response function (which is singular at $2\Delta$) with a Lorentzian 
``damping function'' \cite{yulu}.  This 
however leads to the transfer of rather too much weight into the subgap
region ($\omega < 2\Delta$).   It has been argued that the effect of 
both thermal and quantum fluctuations in the ordered phase can be modeled 
with a zero mean ``white noise'' potential (i.e. one with only delta function 
correlations in real space) \cite{ross}.  Predictions based on this model have 
been sucessfully applied to optical conductivity spectra for the \cdw\ states in KCP 
($K_2Pt(CN)_4Br_{0.3}3H_2O$) \cite{kim} and  Blue Bronze ($KMnO_3$) 
\cite{grunerMcKenzie}.  In the case of \taisy\ \cite{grunerMcKenzie}, while the 
spectra showed some deviations from the predicted behavior, the temperature 
dependance of spectra for $T<\tdot$, $\omega < 2\Delta$ led the authors to 
conculde thermal that lattice fluctautions are indeed an important 
effect, even in the fully ordered \cdw\ state.

\section{Comparison with Strongly Correlated Models}

The properties of (quasi) one--dimensional strongly correlated
electron systems have been studied theoretically for more than twenty years 
and good reviews exist \cite{giamarchi}.
We shall only highlight the transport properties of 
interacting systems which differ radically from
non--interacting ones in ways which are relevant for the experiments
discussed above.

We may characterize the Luttinger Liquid (LL) state
as a) {\it spin--charge separated} -- low energy spin and charge
excitations
are independent of one another; b) {\it gapless} --- low spin and
charge
excitations are bosonic collective modes with ``acoustic'' dispersion
$\omega_{\rho(\sigma )} = v_{\rho(\sigma )} |q|$; and c) critical --
all correlation functions decay with power laws determined by
two parameters interaction $K_{\rho}$ and  $K_{\sigma}$.
(We assume spin rotation symmetry below, in which case $K_{\sigma}
\equiv 1$.  For a recent review of LL physics,
see \cite{johannes}).

These properties are most conveniently illustrated by the asymptotic
form
of the single particle correlation function
\begin{eqnarray}
\label{eqn:llG}
G(x,t) &=&
   \frac{1}{2\pi} \left[
   \frac{e^{ik_fx}}{\sqrt{x-v_{\rho}t} \sqrt{x-v_{\sigma}t}}
   + \frac{e^{-ik_fx}}{\sqrt{x+v_{\rho}t} \sqrt{x+v_{\sigma}t}}
   \right]
   \left[
   \frac{\Lambda^2}{x^2 - v_{\rho}^2t^2}
   \right]^{\theta}
\end{eqnarray}

\noindent
where $x, v_{\rho}t, v_{\sigma}t \gg \Lambda$, a cutoff (interaction
range), and
\begin{eqnarray}
\theta &=& \frac{1}{8}
   \sqrt{K_{\rho} + \frac{1}{K_{\rho}} - 2}
\end{eqnarray}

\noindent
Local repulsive interactions lead to $1/2 < K_{\rho} < 1$ [$0 <
\theta < 1/8\sqrt{2}$].

It is immediately evident from this correlation function that the
Luttinger Liquid does not support Fermionic quasiparticles --- it has
no pole structure in the propagator and therefore no Fermi surface.
That the density of states vanishes as a power law
\be
\label{eqn:DOSLL}
D(\omega) &\sim& \omega^{2\theta},
\en
and that the Fermi surface ``step'' is replaced by a weaker algebraic 
decay $|n(k) - 1/2| \sim |k-k_f|^{2\theta}$ follows from setting $x=0$ 
($t=0$) in \eref{eqn:llG} and Fourier transforming on $t$($x$).
In fact the spectral properties of the models leading to
\eref{eqn:llG} have been worked out in some detail \cite{volker}, and
represent a convolution of the responses of the spin and charge
sectors.  For a general ``universal'' parametrization of the interaction, the
single electron spectral function has two singularities, dispersing as
$v_{\rho}(k-k_f)$ and $v_{\sigma}(k-k_f)$.   The extent to which the spin--charge separated
model is not universal in its spectral properties is dicussed in
\cite{newvolkerpaper}.

Like the free electron gas, the Luttinger liquid is a perfect
conductor, and the finite conductivity of any experimental realization
of a Luttinger liquid would be determined by the effect of additional
terms in the Hamiltonian arising from disorder or coupling to
external fields (crystal field, phonons, etc.).
These are strongly renormalized by interaction \cite{kfc},
and scaling arguments lead to power law temperature dependence of
$\sigma_0$,
and power law behavior in $\omega$ for $\sigma(\omega)$
arising from those (Umklapp) terms which open a gap to charge
excitations \cite{giamarchi, note}.
\begin{eqnarray}
\rho(T) &\sim& T^{2 - \nu}
\label{eqn:LRAsigma0T}\\
\sigma(\omega) &\sim& \omega^{-\nu}
\label{eqn:LLoptical}\\
\nu &=& 4n^2 K_{\rho} - 5
\end{eqnarray}

\noindent
where $n$ is the commensurability of the system (inverse of number of
electrons per site).  \taisy\ is naively a quarter filled system, in
which
case $n=2$ and
\begin{eqnarray}
\nu&=& 16 K_{\rho} - 5
\end{eqnarray}

\noindent
Room temperature optical conductivity and ARPES measurements show
clear
evidence of a gap scale $\Delta$, which means that the conducting
phase of \taisy\ {\it cannot} be a LL by our definitions above.
Nonetheless we might can still look for evidence
of strong correlation effects for $\omega \gg 2\Delta$, with
the understanding that structure at lower energies will be
non--universal and rather more complicated.

Fitting the $\omega^{-3}$ tail of the optical conductivity leads
to the value $K_{\rho} = 1/2$ -- the extreme
limit of what may be accomplished with local electron--electron
interaction, but not incompatible with strong electron--phonon
interaction.  In this case we expect a gap to open in the
spin sector and a renormalization
\begin{eqnarray}
\tilde{K_{\rho}} &=& \sqrt{\frac{m}{m^*}}K_{\rho}
\end{eqnarray}

\noindent
where we have followed the convention usual in CDW
literature of associating a renormalized electron mass
$m^{*}$ with the dynamics of the \cdw.  Empirically this
may be as much as a few hundred times $m$.

Self consistently using the value $K_{\rho} = 1/2$ in
\eref{eqn:LRAsigma0T}
we would then anticipate $\sigma_0\sim 1/T$, which is entirely
incompatible
with the monotonically {\it increasing} conductivity measured
experimentally.
To attempt to extract a value of $K_{\rho}$ by fitting the vanishing
of the density of states at the chemical potential with the form
\ref{eqn:DOSLL} is clearly perverse in a system with so large a gap, and
leads to even more extreme values of $K_{\rho} \approx 1/3$.

It is worthwhile noting that the conductivity calculated above is the
the {\it hydrodynamic} conductivity of the liquid (the response of an
isolated
system in equilibrium to an applied field), and {\it not} an
experimental
conductance.  Predictions for the Landauer (two terminal) conductance
of an
isolated LL depend on how one considers it to be coupled to the
current carrying leads of the experimental apparatus \cite{fisher,
safi}.
As far as we are aware, the question of how a real system of many
chains couples to external current carrying leads has not been
considered
in detail by any author.

The LL is thus clearly a poor candidate for describing
the low energy properties of the conducting phase of (TaSe$_4$)$_2$I:
experiment shows clear
evidence of a pseudogap at room temperature in both spin and charge
channels, and even above this gap
energy scale it does not obey the universal properties
scaling expected of a such a model.

Another paradigm for a one--dimensional
conducting state is the Luther--Emery (LE)
Liquid, which is spin--charge separated and has the
same
charge excitations as a LL, but a gap to spin
excitations.  Moreover, while a LL supports both \cdw\ and spin
density
wave excitations decaying with the same power law, the spin gap
in the LE Liquid singles out \cdw\ fluctuations, making it
a natural candidate for describing \qod\ \cdw\ systems.

The LE Liquid should respond in the same way as the LL
in all responses which couple to charge alone, and so it will have
the same ideal hydrodynamic conductivity as a LL or a free electron
gas.   An important consequence
of this is that its optical conductivity should obey the same power
law behavior as that of a LL.  The spin--gap is felt however, whenever
a probe measures spin excitations (susceptibility measurements) and
also if it couples directly to electrons, as in photoemission,
tunneling
experiments, and two terminal conductances, so all of these should
show evidence of a (quasi--)gap.  As both charge and spin sensitive
probes reveal clear evidence of a gap in \taisy, we do not
feel that there are any strong arguments in favour of the LE Liquid
in this case.
That the LE Liquid may provide a rather better description of
Blue Bronze has been suggested by Voit \cite{voitsigma0}.

\section{Core Hole Spectra}

Core level X-Ray photoemission spectroscopy (XPS) provides a
useful probe of the structure of a material, and of the low energy
excitations of any charge carriers present in it.
Generically these spectra have the form of a set of lines in an
incoherent background, at energies many electron volts below the 
chemical potential.  Because the wavefunctions of the electronic core 
levels are physically very small, they provide an essentially {\it local} 
probe of the structure of the material, with shifts in individual 
lines providing information about changes in the electric field and 
charge susceptibility at the sites of individual atoms.

The details of the lineshape of electrons emitted from
core levels also depend sensitively on the {\it low energy}
excitations of any free charge.  These
excitations are in turn dominated by many--body effects and the
asymmetries observed in lineshapes for metals are closely related
to another much studied problem -- the ``Fermi Edge Singularity''
(FES) in X--ray absorption.  We will argue that, taken
in conjunction with a complementary local probe of low energy
spin excitations such as NMR, XPS (or, with some modification to our
arguments, the FES) offers a novel means of
distinguishing between different metallic and conducting phases.

The calculation of many body effects in core level and X--ray edge
response in three dimensions for a Fermi Liquid has been
extensively studied in a body of work commonly referred to as MND
(Mahan--Nozi\`eres--Dominicis) theory \cite{standardrefs}.

In the case of ordinary metals
it is enough to consider the noninteracting problem.
\begin{eqnarray}
H &=& H_0 + V\\
H_0 &=& \sum_k \epsilon(k) c^{\dagger}_k c_k + \epsilon_h
d^{\dagger}d\\
V &=& \frac{1}{L} \sum_{k,q} V_q c^{\dagger}_{k+q} c_k dd^{\dagger}
\end{eqnarray}

\noindent
where $c^{\dagger}_k$ creates a conduction electron with energy
$\epsilon(k)$
and $d^{\dagger}$ creates a core electron with binding energy
$\epsilon_h$.
In higher dimensions $\{k,q\}$ are vectors $\{\vec{k}, \vec{q}\}$.
Spin indices have been suppressed for compactness.
We make the usual assumption that lineshape for the core level
measured
in an XPS experiment is proportional to the Fourier Transform of the
(retarded)
core hole Green's function
\begin{eqnarray}
G_h(t) &=&  -i \theta(t) \langle \mid d^{\dagger}(t) d\mid \rangle
\end{eqnarray}

\noindent
broadened by an appropriate factor to allow for experimental
resolution.  In practise to extract a usable lineshape allowance must
also be made for the finite lifetime of the core hole (modeled as a
Lorentzian
with width set by Auger decay processes) and in some cases a symmetric
(gaussian)
broadening of the line due to interaction with phonons.  These issues
are
discussed in the associated literature \cite{doniach}.

An important many body effect comes into play in the evaluation of
this correlation function.  As originally observed by Anderson
\cite{anderson}, the
ground states of a free electron system with and without the impurity
are
orthogonal, their overlap vanishing as $1/N^{\alpha_{0}}$, where
\begin{eqnarray}
\alpha_0 = \frac{1}{2} \sum_l \left( \frac{\delta_l}{\pi} \right)^2
\end{eqnarray}

\noindent
$N$ is the total number of particles and $\delta_l$ is the phase shift
in the $l$-th scattering channel.
In three dimensions $l$ has the natural interpretation of an angular
momentum quantum
number for scattering states.

In metallic systems this orthogonality makes itself
felt in XPS line asymmetries and in a contribution to the FES
in X--Ray absorption; the number
of zero energy
particle hole pairs which may be made at the Fermi edge is essentially
only bounded by the level spacing of the system, which in the
thermodynamic limit, for a $d$--dimensional system in volume $L^d$
vanishes as $1/L$.
An important consequence of this for XPS is that in removing a core
electron
from a metal, a very large number of conduction
band electron--hole pairs are excited by the unscreened core hole.
This converts the lineshape measured by and \xps\ experiment from a
simple
$\delta$ functional form
\begin{eqnarray}
G_h(\omega) &\propto& \delta (\omega - \omega_T)
\end{eqnarray}

\noindent
to a power law asymmetry
\begin{eqnarray}
G_h(\omega) &\propto& \frac{1}{ (\omega - \omega_T^{\prime})^{1 -
\alpha} }
\end{eqnarray}

\noindent
where $\omega_T$ and $\omega_T^{\prime}$ are the threshold energies
for
emission
of a core electron (naively given by $\epsilon_h$).

For free electrons the XPS exponent $\alpha$
differs simply by a factor of two from the orthogonality exponent
for the overlap $\alpha_0$ defined by Anderson.
Within the Born approximation,
$\delta_l$ 
will be proportional to the modulus of the matrix element for the 
core hole potential in that channel: $\delta_l \approx |V_l|/v_f$.
In one dimension there are only two
scattering channels ``forward'' ($q \approx 0$) and ``backward''
($q \approx 2k_f$), so,
\begin{eqnarray}
\label{eqn:born}
\alpha ^{\mbox{\scriptsize BA}} = n_0^2 (|V_0|^2 + |V_{2kf}|^2)
\end{eqnarray}

\noindent
where $n_0=1/2\pi v_f$ is the density of states at the Fermi surface.

X--Ray spectroscopy of this type is a probe of the
low energy {\it charge} excitations at a given lattice site, 
and as such essentially  measures its local dielectric constant.
We can infact generalize the result Eq.\ \eref{eqn:born} to an arbitrary 
many electron system, using the relation 
\be
\alpha &=& \lim_{\omega\to 0}\sum_{q} \mid V_q \mid^2
  \frac{\Im\{\chi_{\rho}(q,\omega)\}}{\omega}
\en
which we derive in Appendix C, {\it provided} that 
certain conditions on $\chi_{\rho}(q,\omega)$ and $V_q$ hold.

There is also a closely related shift in the XPS line position
\be
\Delta E &=& \sum_{q} \mid V_q \mid^2
  \Re\{\chi_{\rho}(q,0)\}
\en
Both of these parameters acquire temperature dependence in a system
where the charge susceptibility changes with temperature.

Returning to the case in point --- \taisy --- since the density of states at the
Fermi energy is zero in a mean field charge density wave state, and strongly
suppressed for an LRA Liquid  Eq.\ \eref{eqn:dosef}, we clearly expect very little
asymmetry in XPS spectra for \taisy\ in either its metallic or conducting phase.  
Such asymmetry as exists would follow the form in \Fref{fig:shift}.

The temperature evolution of $\Delta E(T)$ may well be measurable.
However since processes at all energy scales contribute to the 
real part of the local susceptibility, and the absolute value of the
shift is cutoff-dependent there is no simple universal formula for the 
shift valid at all teperatures.  The temperature 
dependance of a relative line shift must be found numerically.
Results for this quantity are plotted for our parameterization of the model 
in \Fref{fig:shift}.

This lack of asymmetry is to be expected: 
many body effects related to the existence of a 
Fermi surface will not occur in a system with a gap (or quasigap).
However, such an argument can only be applied in a system where the
quasi--particles are in some sense electron---like.  It is clearly
unreliable in the case of a Luttinger Liquid where the quasi--particles are
stable {\it but} are not electron--like in character.  We consider this
case below.

The X--ray response of strongly correlated one dimensional metal
(a Luttinger Liquid) has become something of a {\it cause
c\'{e}l\`{e}bre}
in recent years.  ``Forward scattering'' from a core level may easily
be treated within the bosonization scheme used to diagonalize
the interacting problem, leading to a suppression (enhancement)
of the asymmetry measured in a system with repulsive (attractive)
interaction between electrons, compared with that measured for
non--interacting electrons \cite{alpha}.

In \ref{XPSinLL} we present an exact solution of the forward
scattering contribution to XPS in a Luttinger liquid; here we simply
reproduce the result in convenient form
\begin{eqnarray}
\label{eqn:LLalpha}
\alpha_f^{\mbox{\scriptsize BA}}(V)
=K_{\rho} \frac{v_f^2}{v_{\rho}^2} \alpha_f^{\mbox{\scriptsize BA}}
\end{eqnarray}

\noindent
where $v_{\rho}$ and $K_{\rho}$ are the parameters of the
charge sector of the LL, defined in \Eref{eqn:llG}, and 
$\alpha_f^{\mbox{\scriptsize BA}}$ is the result for non
interacting electrons within the Born Approximation.

Unfortunately ``backward scattering''
introduces terms nonlinear in bose fields into the Hamiltonian and
make
the problem very much harder to treat analytically.   Renormalization
group analysis in fact suggests that strength of this nonlinear term
flows to infinity, ``breaking'' the one--dimensional chain of atoms
\cite{kfc}.  Within this ``open chain'' interpretation of the RG
results, the contribution to the orthogonality
exponent
$\alpha_O$ from backward scattering is expected to take on the
universal
value of $1/16$, and the orthogonality exponent
has the value
\begin{eqnarray}
\alpha_O = \frac{1}{2} \left(\frac{\delta_f}{\pi}\right)^2 +
\frac{1}{16}.
\end{eqnarray}

\noindent
This result may be derived very elegantly from the use of boundary
conformal
field theory \cite{affleck}.

Equivalent results for XPS response have been derived in expansions
about the ``open chain'' fixed point \cite{analytic}.  Generically
these
predict
an exponent
\begin{eqnarray}
\alpha_E = 2K_{\rho} \frac{v_f^2}{v_{\rho}^2}
\alpha_f^{\mbox{\scriptsize
BA}} +
\frac{1}{8}
\end{eqnarray}

\noindent
at threshold, with a crossover to a different non--universal
scaling law at a finite energy determined by the strength of the
core--hole conduction electron coupling.   A particularly elegant
approach to the XPS problem which reproduces these and many other
standard results is provided by the use of boundary conformal
field theory \cite{affleck}.

It should also be noted that not all authors have found their results
in agreement with the ``open chain'' interpretation set out above,
with one asserting that backward scattering terms can lead to an
{\it enhancement} of the FES \cite{of}.  A recent numerical treatment
of the problem with references to earlier analytical work is provided by
\cite{dmrg}.

No predictions have yet been made for the temperature dependence of
the XPS exponent in a LL, but we can speculate
that these will also obey non--integer power laws controlled by
$K_{\rho}$.
As stated above XPS measures the leading $\omega$ 
dependence of the local charge susceptibility a system.
For a LL, we may divide the susceptibility into a regular part
at $\chi_{\rho}(q\approx 0)$ and an anomalous part at
$\tilde{\chi}_{\rho}(q\approx 2k_f)$.  The former has the same
structure as in a free electron gas (up to charge velocities); from scaling
arguments we anticipate that the temperature dependence of the latter
will be a power law controlled by the interaction parameter
$K_{\rho}$.

As stressed above, XPS is a zero energy probe and the scaling
of $\tilde{\chi}_{\rho}$ will
be cut off by any gap.

Core level studies for materials with a \cdw\ groundstate
(see, \eg , Ref.\ \cite{franz}) usually reveal gross differences between spectra 
taken above and below the transition temperature.  There are two fundamental 
reasons for these differences -- the change in the local environment of core
states due to a structural transition or \cdw, and changes in
the charge suspcetibilty induceded by the \cdw\ transition, which modify the 
lineshape and position, as discussed above.

In a one--dimensional tetramerized charge density wave state
(assuming perfect commensurability) there are two inequivalent sites 
in each unit cell.  The unbalanced charge accumulated due to the charge
density wave will be different at each of these sites leading to a 
splitting in the XPS lines proportional to
the magnitude of the charge density wave $\langle \rho(2k_f,T) \rangle$,
and therefore following it in temperature dependence.
It is unlikely that such a splitting would be observed in 
an incommensurate \cdw\ system like \taisy\, although this
has been a subject of some contention \cite{sato}.

Incommensurability can be expected to lead to the smearing of the
core level threshold energy over a range of energies again proportional to the
magnitude of the charge density wave, and therefore to lines whose width 
follows $\langle \rho(2k_f, T) \rangle$.   In
the quasi--static picture on which our model is based on the splitting
(or smearing) of the core line observed in the ordered phase would
persist into the conducting phase, since the structural deformations
of the lattice persist --- the loss of order is a
loss of three--dimensional coherence only and is essentially
irrelevant to the site--local physics of core holes.

Importantly however, a splitting (smearing) of the line due to the
inequivalence of $Ta$ sites will not alter the asymmetry of the
underlying lineshape.  {\it If} there are zero energy charge
excitations in the conducting phase the measured split (smeared)
lineshape should also be asymmetric.   Shifts in the line 
due to changes in the local susceptibility will persist for
the same reason.  In bulk $Ta$ the $4f$
orbitals may be fitted with a Doniach--Sunji\'{c} lineshape and
has an asymmetry $\alpha_{E} \sim 0.1$.

There are a number of other effects which need to be borne in mind.
Foremost among these is spin--orbit coupling which can
split core lines by large energies.   In the case of
the $Ta_{4f}$ state discussed above spin--orbit coupling leads to
a splitting of $\sim 3 eV$ between $4f_{7/2}$ and $4f_{5/2}$ states.

We should also be aware that an attractive potential (such as a core
hole) in a one dimensional electron system will generally have a
bound state.
The presence of this bound state will modify the threshold for
photoemission out of the core level (giving in general two
thresholds, one
for sites with the bound state filled and one for those where it is
empty) and, in a truly metallic system, will also modify the asymmetry
measured in each of these cases \cite{nozieres}.  As splitting due to
bound states will be uncorrelated with the lattice distortion, it can
in principle be distinguished from one due to inequivalent
sites in a distorted lattice.  The possibility of a line split by a
bound
state appearing as a single (asymmetric) line in XPS carried out at
finite resolution poses a more serious threat to the interpretation
of data
on these systems, and would need to be considered before very strong
conclusions were drawn about any measured asymmetry.

\section{NMR}

NMR provides a local probe of spin excitations which is exactly
complementary to XPS (a probe of local charge excitations).  In fact,
up to a structure factor $F(q)$, the nuclear relaxation rate $1/T_1$
is given by precisely the same formula as we derived in \ref{xps} for 
the XPS asymmetry, but with the {\it spin} susceptibility $\chi_{\sigma}$ 
substituted for the {\it charge} susceptibility $\chi_{\rho}$ 
and the structure factor $F(q)$ substituted for the core--hole 
conduction electron interaction matrix element $\mid V_q \mid^2$.
\begin{equation}
\frac{1}{T_1 k_B T} = \frac{\gamma^2 }{2 \mu_B^2}
\lim_{\omega \rightarrow 0} \sum_q F(q)
   \frac{\Im \{\chi_{\sigma}(q, \omega)\}}{\omega}.
\end{equation}
Here $\mu_B$ is the Bohr magneton and the nuclear spin has
gyromagnetic ratio $\gamma$.  We shall assume purely on-site coupling.  
This leads to a momentum-independent coupling $F$.  Then the rate is simply
proportional to the slope of the absorptive part of the local spin
susceptibility at very low frequencies.  We shall neglect spin-orbit
coupling so that the spin susceptibility is well-defined and
isotropic.

Since the single-particle Green's function for the LRA liquid is diagonal in spin 
indices, we may evaluate the spin susceptibility in terms of the 
density of states in the same way as we evaluate the charge suscpetibility 
in \ref{xps} to find 
\be                                                                                              
        \frac{1}{T_1 k_B T}     &=& 2\pi\gamma^2 k_B F n_0^2.                                                    
\en
where $n_0$ is the density of states at the Fermi energy.\
For a regular metal (Fermi Liquid) where $n_0$ is not a function
of temperature this gives a constant relaxation rate (Korringa Law).

The quantity $n_0^2$ is however very strongly temperature dependent in 
the LRA theory (\Fref{fig:n0sq}) and so the temperature dependences of the relaxation 
rate and can be used directly to measure of the extent to which 
the pseudogap depresses the density of states at the Fermi energy. 
NMR measurements are in principle possible for \taisy\ since $Se$ at 
least is NMR active, but no data at present exist.

The NMR Knight shift is determined not by the local, but by the uniform
susceptibility 
\be                                                                                              
\frac{\Delta H}{H_0} &\sim& \chi_{\sigma}                                
\en
and so follows this in temperature dependence.   This is plotted in \Fref{fig:chi}
for our paramatrization of the LRA model.  

In a LL, the temperature dependance of $1/T_1T$ is dominated at low 
temperatures by the scaling dimension of the $q \sim 2k_f$ part of the spin 
susceptibility, which is set by the anomalous dimension of the charge 
field $K_{\rho}$
\be                                                                                              
        \frac{1}{T_1 T} &\sim&  1 
        + \left(\frac{T_0}{T}\right)^{1 - K_{\rho}}.                                                
\en
where $T_0$ is the characteristic temperature at which the $q\sim 0$
and $q \sim 2k_f$ susceptibilities are of equal size.

Data for the \qod\ organic spin density wave (SDW) system $(TMTSF)_2 CI0_4$
have been interprited as showing LL behaviour over a range of 
temperatures just above the SDW transition temperature, with the rather 
small value of $K_{\rho} \sim 0.15$ \cite{nmrLL} found from a fit to $1/T_1T$.   
As far as we are aware no such analysis of NMR data has been attempted for a 
CDW system.

\section{Comparison with other Theories}

We have presented predictions for XPS and NMR.  This has a particular 
goal.  In a theory such as the LRA model the pseudogap is felt
equally in both spin and charge channels, and should be evident in
the suppression of both XPS asymmetry and nuclear spin relaxation.
Importantly, both are {\it local} probes of
the charge(spin) susceptibility and as such, directly comparable.

In the limit of large coherence length the LRA spectral 
function \eref{eqn:A} reduces to the sum of two lorentzians.
In the language of many--body Green's functions we may say that the
poles associated with fermionic quasi--particle excitations have migrated
from the real axis into the complex plane, becoming purely 
imaginary for zero frequency.  The fact there are still simple 
poles in the electron Green's function 
means that these quasiparticle excitations are
electron--like in nature and therefore carry both spin and charge.  
If the effects of the mean field gap persist in the
spin channel for the metallic phase, they will also persist in the
charge channel, and the (quasi--)gaps measured in each channel should
have the same temperature dependence  --- in the language associated 
with high-temperature superconductivity, the ``quasi--gap'' and the
``spin--gap'' should be in exact correspondence.

We can make this statement more formal by considering
the ratios of the XPS assymetry and the NMR relaxation time
\be
R &=& \frac{\alpha(T)}{1/T_1 T}
\label{eqn:ratio}
\en 
Within the LRA model (or any other model in which the pseudogap is felt 
equally in spin and charge channels) $R$ should be essentially 
temperature--independent.  

Almost none of these features are shared with the canonical
one--dimensional model of interacting electrons which we may
most meaningfully compare with experiments on \taisy\ (and other
\qod\ \cdw systems)  --- the Luther-Emery (LE) model.
Here there are no fermionic quasiparticles and the gap exists uniquely
in the spin excitations of the system.
This means that the ratio $R$ of \Eref{eqn:ratio} should be very 
strongly temperature dependent, due to the activated behavior 
of spin excitations.

A plausible form for the spectral function of the LE liquid
capturing the essential features of the gap in spin
excitations the has been proposed by Voit \cite{johannesLE}.
The spectral function has two dispersing
maxima, one associated with charge and the other with spin
excitations, but now only the spin peak is a true singularity;
the divergence at the charge peak is cut off by the spin gap.

The spectral properties of a related field--theoretical model of
electrons interacting with phonons in one--dimension have also been
derived by Wiegmann \cite{paul}.   This goes somewhat beyond the considerations
of the other theories alluded to in this article in identifying the
readjustment
of the order parameter which accompanies the introduction or removal
of an individual electon.   It is characterized by {\it extremely} 
broad and {\it asymmetric} dispersing features, even in the absence of
electron--electron interaction.  The extreme breadth of the dispersing 
features in this model resembles photoemission spectra taken on the
high-T$_c$ cuprates rather more than those taken on quasi one--dimensional
conductors.

Because of the broadness and asymmetry of features in the LE spectral
functions and its dual peak structure, it
would be difficult to extract a single energy scale from it empirically.
An experimental spectrum taken at finite resolution might well ``measure'' 
a different gap
scale and temperature dependence different from that determined by
transport, optical conductivity, or susceptibility experiments.   

Thus the LRA model
offers a better description of existing photoemission
data {\it for this system} \cite{shannon}.  The LE model may prove to be
more nearly applicable to the other most commonly studied inorganic
\qod\ \cdw\ system, Blue Bronze ($K Mn O_3$) \cite{voitsigma0}, but
faces some challenge in explaining the suppression of charge
excitations over the energy range $0.1 \to 0.3eV$ seen in optical
conductivity.

\section{The gap discrepancy}

There remains the discrepancy in gap values mentioned at the outset.

First note that only the photoemission value is truly out of line.
Based on this we may put forward three speculative resolutions.

One is that photoemission,
which takes place on a very fast time scale and is a high energy
process which may couple to many excitations besides the formation of
a hole,  cannot be compared directly with a long--time scale, low energy
experiment such as transport.  This point of view is broadly compatible with the
treatment of the electron--phonon problem performed by Brazovskii and 
Dzyaloshinskii \cite{brazovskii} in which there is a clear separation of time scales
between different electron and phonon excitations of the system.

Second, a movement of intensity away from the Fermi energy in 
photoemission spectra could originate in ``extrinsic'' energy loss
processes,
\ie , those affecting the outgoing electron.  In the range considered,
Drude and phonon losses are the most important loss mechanisms.  
These extrinsic processes may be independently measured by 
electron loss spectroscopy \cite{me}.

Third, midgap states may explain the mismatch between photoemission
and transport gaps.  These are highly localized states
in the middle of the gap, which cannot carry a current and (due to
their small number) have a small cross
section for photoemission, but which offer a reservoir of charge
that can be thermally excited over (half) the gap and participate
in transport.   This conjecture leads directly to the prediction that
the activation energy measured at low temperatures is of order $\Delta$
and not $2\Delta$.  The transport activation energy is indeed approximately
half the gap measured in photoemission, and a possible mechanism for the
generation of localized mid--gap states does exist in \taisy\ through
interaction with the lattice distortion.
There are four equivalent ways of accomplishing the nearly
commensurate
tetramerization of the lattice which is observed experimentally, and
it
is natural to suppose that this ``frozen'' phonon is divided into
domains with
different phase by kinks (solitons) in the order parameter.
That kinks in an order parameter can
bind (fractional) charge is well known and has been extensively
studied in the case of polyacetylene \cite{ssh}, which has a half
filled
band.   A numerical study of the formation of mid--gap states in the
nearly
commensurate quarter--filled case was performed by Machida and Nakano
\cite{midgap},
who find fractal structure in the density of states reminiscent of the
Hofstader
problem \cite{hofstader}; in fact both the motion of an
electron in an incommensurate potential in one dimension and
in a magnetic field at incommensurate Landau level can be described
by the same {\it almost Mathieu} equation.
As we
have been arguing that the lattice distortion persists in the normal
state of \taisy\ (a weak superlattice corresponding to the
condensed TA Phonon mode is still visible in X--Ray experiments at
300K)
we can further speculate that the midgap states will persist to these
temperatures, leading to the observation of the smaller ``gap'' in
room temperature optical conductivity.  However,  this hypothesis
has the weakness that the states involved must have sufficient weight to
dominate transport properties but not enough to show up in photoemission.

Note that tunneling experiments on \taisy\ 
would be extremely informative.  This experiment also 
measures the density of 
states.  Given that photoemission is in conflict with other data, a 
check is necessary.

Thus the different gap sizes measured by different probes in quasi--one
dimensional systems are hard to reconcile with the LRA model unless
some additional mechanism is postulated {\it a posteri} to explain the
discrepancy.
They may be somewhat less surprising within a strong interaction
scenario, but the different
determination of (pseudo--)gap sizes and
temperature dependence by different probes is {\it not} a problem
of the exotic non--Fermi liquid conducting phase only, but one which
needs to be addressed also in the context of the ordered phase.
In the metallic phase however, XPS provides, at least in
principle, a novel means of exploring whether (quasi--)gaps exist for
both spin and charge excitations or only for spin, and might be
particularly
informative if taken in conjunction with NMR experiments.

\section{Comparison with Underdoped High-temperature Superconductors}

What light does \taisy\ shed on the pseudogap in high-T$_c$
superconductors ?  Photoemission results on the two systems are the
obvious place to start.  In underdoped Bi$_2$Sr$_2$CaCu$_2$O$_{8+x}$ above
T$_c$ the peak of the spectral function $A(\vec{k}, \omega)$ never reaches
the Fermi energy, stopping about $\sim 10 meV$ below the Fermi energy when
$\vec{k}$ is along the $(0,\pi)$ direction \cite{loeser}.  Ths
precise size of this pseudogap depends on doping and temperature and 
vanishes at a temperature $T^*>T_c$.   This observation is qualitatively 
similar to that in \taisy.  There is, however, more than an order of
magnitude difference in the size of the pseudogaps.  Furthermore, the
pseudogap
in  Bi$_2$Sr$_2$CaCu$_2$O$_{8+x}$ is anisotropic with little or no
pseudogap
observed in the $(\pi,\pi)$ direction.  This possibility does not
arise in \taisy.  The similarities
go beyond this.  The uniform magnetic susceptibility is suppressed in
underdoped high-T$_c$
systems \cite{liang} and in \taisy\ \cite{johnston1}, showing that
the pseudogap is present
in the spin channel as well.  The temperature dependence of the
resistivities $\rho(T)$ is
superficially very different.  In the underdoped high-T$_c$ systems,
$\rho$ rises rapidly
and monotonically between T$_c$ and $T^*$,
though there are strong deviations from the linear behavior of
$\rho(T)$ which is characteristic of the optimally doped material.
In \taisy, $\rho$ is roughly independent of temperature in
experiments, as we have seen.
However, this may only reflect the fact that the accessible regime of
temperatures does not
reach up to $T_{MF}$.  We would expect $\rho$ to rise when $T \sim
T_{MF}$,
as seen in Fig.\ \ref{fig:sigmazero}.  Finally, we may compare optical conductivities
$\sigma(\omega)$.
The most interesting high-T$_c$ data from our point of view are
actually c-axis
conductivities which have no Drude peak at low frequencies
\cite{uchida}.  These show a very clear pseudogap
opening up as a function of temperature.  In the a-b plane
conductivity there is a Drude peak.
This interacts in a subtle way with the suppression of the density of
states and the evolution of the
electron lifetime due to the pseudogap, rendering the interpretation
of the data somewhat complicated.
In \taisy\, we are more fortunate in that the Drude width is
considerably less than the pseudogap energy
and the structures in $\sigma(\omega)$ are therefore well separated.
Again, given the difference in parameters,
it appears that the two systems are rather comparable.  NMR
\cite{alloul} and tunneling \cite{renner}
also show evidence for a pseudogap in underdoped high-T$_c$ systems.
These experiments have not been performed in \taisy, but would be
illuminating, as we have already stressed.

The very strong similarity between the two systems suggests a common
origin for the pseudogap
behavior \cite{mckenzieHTc}.  In \taisy, it is clear that the properties of the
pseudogap phase are essentially determined by
the fluctuations remaining from the low-temperature phase.  There is
a very large difference between the
actual critical temperature and the mean-field temperature.  This is
expected in this quasi-one-dimensional material.  In the high-T$_c$
systems, the difference is not so large, as the system is
quasi-two-dimensional.
If we take into account this dissimilarity it appears that in the
intermediate pseudogap region, the
fluctuations from the ordered phase are the determining factor.
Thus, the comparison supports the point of view that the origin of
the pseudogap in is superconducting
fluctuations.  The anisotropy in the pseudogap is
consistent with this as well.

Spin--charge separated theories of the pseudogap regime of 
high-T$_c$ superconductors analgous to the LE liquid have been advanced 
by several authors \cite{HTcspincharge}.  The comparison of XPS and NMR data for 
these systems along the lines which we suggested for a one--dimensional \cdw\ 
system might also shed light on whether the (pseudo--)gap exists in both 
spin and charge channels.
We hope to develop this idea further in a later paper.

\section{Conclusions}

The \lra\  model offers the simplest possible scenario for
fluctuations of \cdw\  order in \qod\  \cdw\  systems that 
reduces simply to the mean field ordered state.
The naive application of this model to existing data for the \qod\
\cdw\ system \taisy\ is quite successful.  In those cases where the 
LRA model fails to explain experimental data for this system above the 
\cdw\ transition temperature ($T_{3D}$), the usual mean field picture of 
the ordered state must also be called into question below $T_{3D}$.   
The most important of these 
discrepancies is the size of the \cdw\ gap (pseudogap) which is 
found to be of different size by different experimental techniques, 
both above and below $T_{3D}$.

This simple view of the origin of 
a pseudogap in a quasi--one--dimensional system should be contrasted with the canonical picture of 
``gapping'' in 1--D, strongly correlated systems with strong CDW fluctuations, in
which non--current conserving terms in the Hamiltonian lead to the formation 
of a gap for spin, but not
for charge excitations.  In this scenario, dispersing features in the
electronic spectral function are generally extremely broad and
asymmetric, and while the ordered state of the system
must be three-dimensional, it need not be mean field like in nature.
In those cases where we have been able to compare the LRA model directly to
experiments on \taisy\, it seems to offer a better description 
of data than existing predictions for strongly correlated models.

XPS provides, at least in principle, a novel means of exploring which
of these scenarios is correct, and might be particularly 
informative if taken in conjunction with NMR experiments.
Tunneling measurements of the density of states would serve as a useful 
check on photoemission.

\ack

This work was supported under NSF Grant No.  DMR-9704972.
It is our pleasure
to acknowledge informative and helpful converstaions with 
Ian Affleck, Nicholas
d'Ambrumenil, Andrey Chubukov, Marco Grioni, Franz Himpsel, 
Peter Koepitz, Volker Meden, Hartmut Monien, 
Andr\'{e}-Marie Tremblay, Johannes Voit and Paul Wiegmann.

\Figures
\begin{figure}
\caption{a) Structue of $Ta$ chains in (TaSe$_4$)$_2$I.
b) Schematic displacement of $Ta$ atoms in tetramerized ground
state.}
\label{fig:taisy}
\end{figure}

\begin{figure}
\caption{Example of ARPES spectrum for dispersing feature with k=k$_F$
         and theoretical fit, taken from Ref.\ 4.}
\label{fig:edc}
\end{figure}

\begin{figure}
\caption{The temperature dependence of the uniform susceptibility
as a function of temperature.  The points are data and the line is
the LRA model fit, both from Ref.\ 5.}
\label{fig:chi}
\end{figure}

\begin{figure}
\caption{The DC resistivity over the
    experimentally accessible temperature range.  The data in (a) are taken from
    Ref. 12 and the theoretical curve in (b) is plotted using
    Eq.\ 8.}
\label{fig:sigmazero}
\end{figure}

\begin{figure}
\caption{a) Experimental spectra for the optical conductivity 
         (TaSe$_4$)$_2$I taken from Ref.\ 8.
         b) The intrinsic optical conductivity of an LRA liquid as calculated
         from current--current correlations according to Eq.\ 12.}
\label{fig:optical}
\end{figure}

\begin{figure}
\caption{The temperature dependence of the density of states at the
   Fermi energy {\it squared} normalized to free electron values,
   for the LRA model, as parametrized to describe optical
   conductivity measurements.   The temperature dependence of NMR $1/T_1$ and the
   XPS asymmetry $\alpha_E$, should both follow up to the accuracy of this 
   parametrization of the model.}
\label{fig:n0sq}
\end{figure}

\begin{figure}
\caption{Numerically determined change in the local static 
susceptibility $\sum_q \chi (q,0,T)$ relative to 
$T_{\protect\mbox{\protect\tiny 3D}}$.
This determines the shift in XPS lines.}
\label{fig:shift}
\end{figure}

\begin{figure}
\caption{a) MF theory of the charge density wave state in terms of 
multiple scatterings 
between right and left moving electron states.   These may be resummed 
to give a self energy correction which {\it coincidentally} is  
equivalent to b) the first term in a perturbation theory developed
using the $\omega \to 0$ limit of the phonon propagator 
$D(q) \sim \delta_{q\pm 2 k_f} \delta(\omega)$.}
\label{fig:MFvertices}
\end{figure}

\begin{figure}
\caption{Diagrams which must be evaluated when using the $\omega \to 0$ 
limit of the phonon propagator, as in the theories due to Sadovskii.   
Propagators may be modified to take 
account of the finite correlation length of the lattice distortion, 
but the theory is then no longer exactly soluble.   Crossing diagrams 
at higher order {\it are} included.}
\label{fig:sadovskii}
\end{figure}

\begin{figure}
\caption{Diagrams calculated in the ``static non--crossing `` approximation 
with the phonon propagator is $\delta_{q\pm 2 k_f} \delta(\omega)$.}
\label{fig:noncr}
\end{figure}

\clearpage

\appendix

\section{Formal standing of the LRA theory \label{whatislra}}

In this appendix we discuss the nature and validity of the 
LRA theory from a technical rather than 
phenomenological point of view. 

We begin by rederiving the self energy correction \eref{eqn:selfeng}
using equations of motion in the spirit of the original LRA paper, 
but showing that on a more careful examination of the nature
of the averaging involved, the equations of motion do in fact close.

Our starting point is the Hamiltonian 
\begin{eqnarray}
\label{eqn:Heqnmtn}
H_{LRA} &=& \sum_{k} \epsilon (k) c^{\dagger}_k c_k
  +  \sum_{Q, k'>0}
   [ \Psi^*_{-Q} c^{\dagger}_{k'-Q} c^{}_{k'}
   + \Psi_{Q} c^{\dagger}_{-k'+Q} c^{}_{-k'} ]  \\
   \Psi_Q &= & \frac{1}{\sqrt{L}} g(Q) \langle u(Q) \rangle.
\end{eqnarray}
and 

The imaginary time dependence of the single electron 
Green's operator (in the Heisenberg picture) is given by
\be
\hat{G}(k^{\prime}, k, \tau) &=& 
   - T_{\tau} c_{k^{\prime}}(\tau) c^{\dagger}_k (0) \\
   &=& 
   - \{\Theta(\tau) c_{k^{\prime}}(\tau) c^{\dagger}_k (0) 
   - \Theta(-\tau) c^{\dagger}_k (0) c_{k^{\prime}}(\tau)\}
\en
and its evolution follows
\be
\frac{\partial\hat{G}(k^{\prime}, k, \tau)}{\partial \tau} 
   &=& - \delta(\tau) \{ c_{k^{\prime}}(0), c^{\dagger}_k (0) \}
    -  T_{\tau} \frac{\partial{c_{k^{\prime}}(\tau)}}{\partial \tau}
       c^{\dagger}_k (0) .
\en
The Fermi field in turn has dynamics governed by
\be 
\label{eqn:c}
c_{k^{\prime}}(\tau) &=& e^{H\tau}c_{k^{\prime}} e^{-H\tau}\\
\frac{\partial{c_{k^{\prime}}(\tau)}}{\partial \tau} &=& 
  e^{H\tau} [H, c_{k^{\prime}}] e^{-H\tau}
\en
and the phonon field, by explicit assumption, is static 
--- \ie has no dynamics.

Inserting the Hamiltonian \eref{eqn:Heqnmtn} in \eref{eqn:c} we find
\be 
\frac{\partial\hat{G}(k^{\prime}, k, \tau)}{\partial \tau} &=& 
   -\delta(\tau)\delta_{kk^{\prime}}
   -\epsilon(k^{\prime}){G}(k^{\prime}, k, \tau)\\
   && - \left\{\sum_{Q} \Psi^{*}_{Q} \hat{G}(k^{\prime} + Q, k, \tau) 
   + \hc \right\} .
\en 
Then for the Green's operator with $k^{\prime} = k $.  
\be 
\frac{\partial\hat{G}(k, \tau)}{\partial \tau} &=& 
      -\delta(\tau)
      -\epsilon(k)\hat{G}(k, \tau)\\
   && - \left\{\sum_{Q} \Psi^{*}_{Q} \hat{G}(k + Q, k, \tau) 
      + \hc \right\}
\en 
We introduce the Fourier transform
\be
\hat{G}(\tau) &=& \frac{1}{\beta} \sum_{i\omega_n} 
      \hat{G}(i\omega_n) e^{-i\omega_n}
\en
Whence
\be 
-i\omega_n\hat{G}(k, i\omega_n) &=& 
      -1  -\epsilon(k)\hat{G}(k, i\omega_n)\\
   && - \left\{\sum_{Q} \Psi^{*}_{Q} \hat{G}(k + Q, k, i\omega_n) 
      + \hc \right\}
\en
Physically
we are interested in scattering between the two Fermi points, \ie 
$k^{\prime} \approx k_f$, $Q \approx 2k_f$, \ie
\be      
[\epsilon(k) - i\omega_n] \hat{G}(k, i\omega_n) &=& 
      1 - \sum_{Q} \Psi_{Q} \hat{G}(k - Q, k, i\omega_n) 
\en

Similarly,
\be
[\epsilon(k-Q) - i\omega_n] \hat{G}(k-Q, k, i\omega_n) &=& 
      - \sum_{Q^{\prime}} \Psi^{*}_{Q^{\prime}} 
      {G}(k + Q^{\prime} - Q, k, i\omega_n) 
\en    
which implies
\be
[\epsilon(k) - i\omega_n] \hat{G}(k, i\omega_n) &=& 
       1 + \sum_{QQ^{\prime}} \frac{\Psi^{*}_{Q^{\prime}} \Psi_{Q} 
       \hat{G}(k + Q^{\prime} - Q, k, i\omega_n)}{\epsilon(k-Q) - i\omega_n}
\en       

We now turn this relationship between operators into an equation for
the single electron Green's function by taking the expectation
value of both sides
\be
[\epsilon(k) - i\omega_n] 
   \langle\langle
      \hat{G}(k, i\omega_n)
   \rangle\rangle
       &=& 1 + \sum_{QQ^{\prime}} \frac{
       \langle\langle 
          \Psi^{*}_{Q^{\prime}} \Psi_{Q} 
       \rangle \rangle
       \langle \langle 
          \hat{G}(k + Q^{\prime} - Q, k, i\omega_n) 
       \rangle \rangle
       }{\epsilon(k-Q) - i\omega_n}
\en    
where $\langle\langle \ldots \rangle\rangle$ denotes both thermal and 
quantum mechanical averageing.

In the case of the classical field $\Psi_{Q}$ quantum averageing is 
irrelevant and we are left with the thermal average
$\langle \Psi^{*}_{Q^{\prime}} \Psi_{Q} \rangle_T$
over all possible configurations of the static phonon field. 
By definition 
$\langle\langle \hat{G}(k, i\omega_n) \rangle\rangle = {G}(k, i\omega_n)$ 
the temperature Green's function for the system.

We note that for a many--chain system with ``frozen''
phonon disorder one might equally motivate an ensemble average on 
potentials from the fact that all experiments measure average over 
many chains, and the potentials on different chains would in general 
be different.

Following Scalapino, Sears and Ferrell (SSF) \cite{ssf}, 
LRA consider consider a one--dimensional 
classical order parameter $\Psi_Q$ described by a free energy
\begin{eqnarray}
F[\Psi_Q] &= & a(T)|\Psi_Q|^2 + b(T)|\Psi_Q|^4 +
c(T)(Q-2k_F)^2|\Psi_Q|^2,
\end{eqnarray}
where the parameters $a(T)$, $b(T)$ and $c(T)$
are taken from the mean field treatment of the electron--phonon
interaction.  Using this free energy to perform the thermodynamic
average it is then found that correlations of lattice at $2k_F$ 
order decay exponentially with distance
\begin{eqnarray}
\label{eqn:lattice}
\langle \Psi(x) \Psi(x^{\prime}) \rangle_T
   &=& \psisq cos[2k_f(x - x^{\prime})] \exp[-|x-x^{\prime}|\invxi]
\end{eqnarray}
where the temperature dependence of the two parameters follows
from $\psisq$ and $\invxi$ may be found from an exact mapping onto
a quantum mechanical anharmonic oscillator problem \cite{ssf}.  

It follows from \eref{eqn:lattice} that
\be
\label{eqn:QQ}
\langle \Psi^{*}_{Q^{\prime}}\Psi_{Q} \rangle_T
      &=& \delta_{QQ^{\prime}}
        \psisq \frac{\invxi}{\pi} 
        \frac{1}{(Q-2k_f)^2 + \xi^{-2}(T)}
\en
and the equations of motion now close.   We assume that within 
this thermal ensemble $\langle \Psi_{Q} \rangle = 0$
which implies that there is no anomalous propagator 
above the three--dimensional ordering temperature $\tdot$.
 
The remaining sum over $Q$ is performed as a contour integral, 
using the result that, for linearized dispersion
\be
\epsilon(k-2k_f) &=& -\epsilon(k)
\en 
yielding 
\be 
{G}(k, i\omega_n)^{-1} &=& i\omega_n - \epsilon(k)
      - \frac{\psisq }{ i\omega_n 
      + \epsilon(k) \pm i v_f \invxi}
\en
with the choice of sign depending on the half plane to which the 
function is to be continued.

The self--energy correction which we derived above
\be
\Sigma_{LRA} (k,i\omega_n) &=& 
   \frac{\psisq }{ i\omega_n 
   + \epsilon(k) \pm iv_f \invxi}
\label{eqn:selfengLRA}
\en
has the correct limits for $\invxi \to 0$ (BCS mean field theory with gap 
$\Delta^2 = \psisq$) and for $\psisq \to 0$ (free electron gas).
Physically it can be thought of as describing the pairing of a 
coherent electron of momentum $k\approx k_f$ with an {\it incoherent} hole 
of momentum $k\approx - k_f$ .  However since it at first sight a somewhat 
crude theory, our use of it requires some clarifcation and justification.

The physical motivation is clear -- below the
mean field transition temperature there is a very strong separation of
time scales between the soft phonon modes and the electrons.
This observation might equally inspire us to develop a
perturbation theory for electron--phonon coupling in which the
electrons scattered elastically from the fluctuations of lattice
order.  This would require 
using the full electron--phonon vertex of the Fr{\"o}hlich Hamiltonian
\eref{eqn:froh} and the $\omega \to 0$ limit of the phonon propagator.

Indeed if we interpret the result of the thermal average over static phonon
fields \ref{eqn:QQ} as an $\omega \to 0$ phonon propagator
\be
\label{eqn:DQ}
D(Q) &=& \psisq \frac{\invxi}{\pi} 
        \frac{1}{(Q-2k_f)^2 + \xi^{-2}}
\en
then the lowest order self energy correction for electrons scattering
off this (quantum mechanical) phonon field is indeed 
the LRA result \eref{eqn:selfengLRA}, and the LRA theory is often discussed
in these terms.

It is not hard to calculate the corrections to this 
self--energy which occur at the next (and higher) order 
in this interaction \cite{shannon1997}.
They are not small, and tend to ``wash out'' the pseudogap
found in the LRA theory --- why then should we trust it ? 

Under the model assumptions made
above about the thermal ensemble of lattice potentials, the equations
of motion for the Hamiltonian \eref{eq:lraH} close.   This means
that the LRA theory we present is the {\it correct} a solution of a slightly 
different problem.  We explain why, and what this problem is, below.

By using the expectation value of the phonon field
in the Hamiltonian \eref{eq:lraH} we are going somewhat further 
than taking the $\omega \to 0$ limit of a quantum mechanical problem;
we are removing the phonons from the quantum mechanics of the problem entirely.
The ``frozen'' lattice is treated strictly as a static external field -- a
classical object. 

This approach has the great advantage of 
reducing exactly to the familiar ``mean field'' case in the 
fully ordered limit where 
\begin{eqnarray}
   \Psi_Q &= & \Delta \delta_{Q\pm2k_F}.
\end{eqnarray}
in which case the Hamiltonian is
\begin{eqnarray}
\label{eqn:MFH}
H &=& \sum_{k} \epsilon (k) c^{\dagger}_k c_k
     + \sum_{k} [\Delta^* c^{\dagger}_{k-2k_F} c_k
        + \Delta c^{\dagger}_{-k+2k_F} c_{-k}]\\
\Delta &=& \frac{1}{\sqrt{L}} g(2_{k_F}) \langle u_{2k_F}\rangle .
\end{eqnarray}

This familiar Hamiltonian may be solved {\it exactly} by the equation of
motion technique used above, or by canonical transformation, and one 
obtains a self--energy of the form
\begin{eqnarray}
\Sigma_{MF} (k, i\omega_n) =
   \frac{\mid \Delta \mid^2}{i\omega_n + \epsilon (k)} 
\label{eqn:BCSselfeng}
\end{eqnarray}

The same self--energy correstion is indeed the result found in a
perturbation theory for electrons interacting with a
$\lim \omega \to 0$ phonon propagator
\begin{eqnarray}
    D(Q) &= & \Delta \delta_{Q\pm2k_f} \delta(\omega)
\end{eqnarray}
but {\it only at lowest order}.
Resumation of the remaining diagrams at higher order \cite{note}
{\it does not} lead to recovery of the correct BCS--like mean field form
for the electron Green's function.  Instead one finds a propagator for a 
system with a {\it distribution of real
gaps}.
\begin{eqnarray}
G(k,i\omega_n) &=& \int_0^{\infty}dx e^{-x}
   \frac{i\omega_n}{i\omega_n^2 - \epsilon_k^2 - x\Delta^2}
\end{eqnarray}

\noindent
a result that first published by Sadovskii \cite{sadovkskii1}.   A later
attempt by Sadovskii \cite{sadovkskii2} to generalize this treatment to the 
case with a finite coherence length (phonon propagator \ref{eqn:DQ}) 
has been widely discussed but seems to be flawed \cite{oleg}.

The reason for the apparent contradiction between the exact result and 
this resumation of perturbation theory is that so far as Hamiltonian
\eref{eqn:MFH} is concerned, the wrong perturbation
theory has been resummed.  The topology of the 
diagrams for scattering off an external field (in our case ``frozen''
phonons) and for scattering from an $\omega \to 0$ phonon propagator
are different because a propagator (two points joined by a line) explicitly 
correlates pairs of scattering events, whilst all scatterings from an external 
field are independant of one another.   

In solving the mean field Hamiltonian we have, in diagrammatic langage, resummed 
the infinite series of all possible mulitiple scaterings which transfer electrons 
between the two Fermi points (see \Eref{fig:MFvertices}).  The vertex in 
\eref{eqn:MFH} is of the form of an external field which scatters particles from 
the right Fermi point to the left (or vice versa) whilst preserving the new momentum $q$.

If one wishes do perturbation theory starting from the Hamiltonian \label{eqn:LRAH}
one should consider diagrams of the same type, for the same reason. 
The coincidence between the first order self energy correction
a perturbation theory developed using an $\omega \to 0$ phonon 
propagator with Lorentzian form and the LRA self energy is in some sense just 
that -- a coincidence -- and moreover the same coincidence as 
the correspondence between the exact self energy of the usual mean field
picture and the lowest order self--energy correction for interaction with 
a phonon propagator of delta function form.  

Our goal is to understand the effects of fluctations of \cdw\ order in 
a quasi--one dimensional system.
In the absence of any exact solution to which we can compare, 
whether it is a better approximation scheme to remove quantum mechanics 
from the phonon field altogether and so be able to recover mean field theory, 
or to consider a problem in which both electrons and phonons are quantum mechanical 
objects but scattering between them is elastic {\it remains an empirical 
question}.   As the thermal averaging over static field configurations
in the LRA scheme in some ways imitates the averaging over many chains
which must take place in the real system, even an exact solution 
which called the LRA scheme into question in 1D might not invalidate its 
application to real systems.

A comprehensive treatment of the Sadovskii model and related technical issues
is supplied by Tchernyshyov \cite{oleg}.  A slightly different 
approach to the same problem has also been considered by Millis and 
Monien \cite{millis}.

\section{Parametrization of Model\label{param}}

The parametrization of the model involves determining the 
temperature dependance of the two parameters of the 
self energy correction, $\psisq$ and the coherence length $\invxi$.
This is to some extent arbritrary.
As we intend to use the model in a phenomenological 
but self--consistent way they should ideally be taken from experiment; in order 
to have a simple working picture we borrow the analysis presented
in the original LRA paper which leans heavily on the results of
SSF \cite{ssf}.

The classical field $\Psi_Q$ has a free energy with parameters 
taken from the mean field (linear response)
perturbative treatment of the 1D Fr\"ohlich Hamiltonian : 
\begin{eqnarray}
\label{freeenergy}
F[\Psi_Q] &= & a(T)|\Psi_Q|^2 + b(T)|\Psi_Q|^4 + c(T)(Q - 
2k_F)^2|\Psi_Q|^2,  
\end{eqnarray}
with
\begin{eqnarray} 
a(T) &= & D_0 \frac{T - T_c}{T},
\nonumber
\end{eqnarray}
\begin{eqnarray} 
b(T) &= & D_0 [b_0 + (b_0 - b_1)\frac{T}{T_c}],
\nonumber
\end{eqnarray}
and
\begin{eqnarray} 
c(T) &= & D_0 \xi_0^2(T), 
\nonumber
\end{eqnarray}
\noindent
where $D_0$ is the (constant) density of states for the band, which is 
taken to have width $2\epsilon_F$.   The parameter $\xi_0(T)$ is the length 
scale emerging naturally from the linear response analysis of the 
one--dimensional electron phonon problem
\be
\xi_0(T) &=& \frac{4\pi k_B T}{\sqrt{7 \zeta(3)} \hbar v_f}
\en
We fix $b_0$ and $b_1$ to give the correct zero 
temperature value of the gap $\Delta_0$ for the experiment under 
consideration $(\Delta_0^{ARPES} \ne \Delta_0^{TRANSPORT})$ according 
to
\be
b_0 =  \frac{1}{2\Delta_0^2},
\en
and
\begin{eqnarray} 
b_1 &= & b_0 \frac{7\zeta(3)}{16 \pi} \frac{(1.76)^2}{0.5}  . 
\nonumber
\end{eqnarray}

The problem of determining $\psisq$ and $\invxi$ then reduces to 
that of finding the low lying energy levels of a particle moving in an 
anharmonic potential well, the shape of which is determined by the 
coefficients of the free energy 
\begin{eqnarray}
\label{anharmonicosc}
H &= & - \frac{1}{4} \frac{k_B^2 T_c^2}{D_0} 
\frac{\partial^2\Psi}{\partial x^2} 
           + a(T)|\Psi|^2 + b(T) |\Psi|^4
\end{eqnarray}

\noindent
This problem may solved numerically, or approximately using perturbation
theory and assymptotic analysis.   

We consider following parametrization, found from a simple 
perturbation theory sufficently accurate in the temperature
range of experimental interest in the next section:
\begin{equation}
\invxi =  \xi_0^{-1}(T) ( \frac{4T}{3T_c} - \frac{1}{3})
\label{xp1}
\end{equation}
\begin{equation}
\psisq =  \frac{a'}{b} (1 - \frac{T}{T_c}) - \frac{1}{2} k_B =
\frac{T_c}{a'}                                            
\frac{1}{\sqrt{1-\frac{T}{T_c}}},
\label{xp2}
\end{equation}
where $a' =  a(T)/T $.
These approximate forms of the parameters $\psisq$ and $\invxi$
are used wherever we compare with expermiment.

One interesting feature of the exact (numerical) determination of the 
$\xi(T)$ is that the correlation length is strictly infinite only at $T=0$
but begins to diverge strongly at a temperature
approximately one quarter of the mean field transition temperature.
This fact leads \cite{lra} to the simple prediction that
\begin{eqnarray}
\tdot \approx \tcmf/4.
\label{eqn:rule}
\end{eqnarray}

\noindent
The value of \tc\ for $\tdot$ found by experiment (Table 1) is
therefore
compatible within a fluctuating \cdw\ scenario with the estimated
$\tcmf$ of $\sim 900K$.

An interesting associated technical question is whether, in
a rigorous treatment of the model defined by \Eref{eqn:froh},
$\xi(T)$ actually does diverge
at zero temperature, or whether it remains finite due to quantum
fluctuations.
This depends on properly including the Umklapp processes present in
a quarter-filled band.  However, for the temperature range $T > T_{3D}$ 
which is of interest in this paper quantum fluctuations are not important.  
For $T < T_{3D}$, three-dimensional effects dominate in the real system,
again suppressing quantum fluctuations.

\def\btt#1{{\tt$\backslash$#1}}

\section{Static noncrossing approximation}

Another possible way of calculating the single-particle Green's function in
the static limit is to replace the phonon propagator
with delta functions:
\begin{equation}
D(q, \omega) = \delta_{q\pm 2k_F} \delta(\omega).
\end{equation}
and in addition to make the noncrossing approximation that
no phonon lines cross.  This amounts to calculating the Green's function
self-consistently but neglecting vertex corrections entirely.
See \Fref{fig:noncr} for an illustration of this.  We shall find that this method
leads to unphysical results.  It is mentioned only because it is one way of
obtaining a pseudogap in a number of different models.

The self energies for two wavectors are
\begin{eqnarray}
\Sigma(k,\omega) &=& \frac{g^2}{\omega - \epsilon_{k - 2k_F} -
\Sigma(k-2k_F,\omega)} \\
\Sigma(k-2k_F,\omega) &=& \frac{g^2}{\omega - \epsilon_{k } -
\Sigma(k,\omega)}.
\end{eqnarray}

Because of the delta-function effective interaction, these equations close.

Consider first the special case $k=k_F$.  Then
$\Sigma(k,\omega)= \Sigma(k-2k_F,\omega)= \Sigma(-k_F,\omega)$,
$ \epsilon_{k} = 0$ and
\begin{equation}
\Sigma(k_F,\omega) = \frac{g^2}{\omega - \Sigma(k_F,\omega)},
\end{equation}
which is a quadratic equation with solution
\begin{equation}
\Sigma(k_F,\omega) = \frac{1}{2} \omega - \frac{1}{2}  \sqrt{\omega^2-4 g^2}.
\end{equation}
The self-energy has a finite imaginary part for $|\omega| < 2|g|$, i.\ e.\ ,
even at zero frequency.

The general case is only slightly more involved.
Let us define $q = k - k_F$ so that
$\epsilon_k = v_F q$ $\epsilon_{k-2k_F} = - v_F q$.
The result is:
\begin{equation}
\Sigma(q,\omega) = \frac{1}{2} (\omega - \epsilon_k) -
\frac{1}{2} \left[(\omega - v_F q)^2 - 4 g^2
\left(\frac{\omega - v_F q}{\omega + v_F q}\right)^2 \right]^{1/2}.
\end{equation}
The imaginary part of this function is nonzero when
$|\omega + v_F q| < 2g$.

A finite relaxation time makes no sense for a
static, ordered, potential.  We conclude that the noncrossing
approximation is not appropriate for this problem.
This is not surprising.  Neglect of vertex corrections
is only valid when the electron wavefunctions are only slightly
perturbed by the external potential.  When degeneracies $(k=k_F)$
or near-degeneracies $(k \approx k_F)$ are a dominant effect, ordinary
nondegenerate perturbation theory is not accurate.

\section{XPS Lineshape and Shift\label{xps}}

We define a core retarded core hole Green's function
\begin{eqnarray}
G_h(t) &=& -i\Theta(t) \langle d^{\dagger}(t) d \rangle
\end{eqnarray}

\noindent
and divide it into two parts
\begin{eqnarray}
G_h(t) &=& -i\Theta(t) e^{-i\omega_T t} \rho(t)
\end{eqnarray}

\noindent
where $\omega_T$ is a threshold energy (containing, in general,
contributions
from all orders of perturbation theory), and all the relevant many
particle
effects
are consigned to the factor
\begin{eqnarray}
\rho(t) &=& \langle \mid T \exp \left[ -i\int_0^t dt_1 V(t_1) \right]
   \mid \rangle
\end{eqnarray}

The lowest order term in a linked cluster expansion for $\rho(t)$ is
\begin{eqnarray}
\label{eqn:begin}
\rho(t) &\approx& e^{F_2(t) }\nonumber\\
F_2(t) &=& \frac{1}{2} (-i)^2
   \int_0^t dt_1 \int_0^t dt_2
   \langle T V(t_1) V(t_2) \rangle\\
       &=& - \frac{1}{2L} \sum_q \mid V_q \mid^2
   \langle T \rho(q, t_1 - t_2) \rho(-q, t_2-t_1) \rangle
\end{eqnarray}
We will confine ourselves here to examining the
contribution to the Fermi Edge Singularity (FES) from this term, 
which contains all the essential physics of the problem.

The expectation value
$\langle T \rho(q, t_1 - t_2) \rho(-q, t_2-t_1) \rangle$, obviously
has the form of a density--density correlation function.  We
therefore define
\begin{eqnarray}
i\Pi(q, i\Omega_m) &=& -\frac{1}{L} \int_0^{\beta} d\tau
   e^{i\Omega_m \tau} \langle T \rho(q, t_1 - t_2) \rho(-q, t_2-t_1)
\rangle
\end{eqnarray}
and introduce a bosonic spectral function $B(q, \omega)$
\begin{eqnarray}
\Pi(q, i\Omega_m) &=&
  \int_{0}^{\infty}
  \frac{d\tilde{\omega}}{2\pi}
  \left[
  \frac{B(q, \tilde{\omega})}{i\Omega_m - \tilde{\omega}}
  + \frac{B(q, -\tilde{\omega})}{i\Omega_m + \tilde{\omega}}
  \right]
\end{eqnarray}
where $i\Omega_m$ is a bosonic Matsubara frequency.

Using the general result
$B(q, \tilde{\omega}) = - B(q, -\tilde{\omega})$, (
which follows from analyticity and parity), we find that
\begin{eqnarray}
\int_{0}^{t} dt_1 \int_{0}^{t} dt_2 \Pi(q, t_1 - t_2)
   &=& 2 \int_0^{\infty} d\tilde{\omega} B(q, \tilde{\omega})
    \frac{1 - e^{-i\tilde{\omega}t}}{\tilde{\omega}^2}
\end{eqnarray}

\noindent
and therefore
\begin{eqnarray}
\label{eqn:theresult}
F_2(t) &=& - \frac{1}{L} \sum_q \mid V_q \mid^2
   \int_0^{\infty} d\tilde{\omega} B(q, \tilde{\omega})
   \frac{1 - e^{-i\tilde{\omega}t}}{\tilde{\omega}^2}
\end{eqnarray}

\noindent
where we have dropped terms linear in $t$ (and therefore
belonging in $\omega_T$).

In order to be able to apply our analysis to a system with
arbitrary spectral function $A(k,\omega)$ we
rewrite the correlation function $\Pi(q, i\Omega_m)$
in terms of Fermionic propagators, and then
substitute a spectral representation of the electrons
\begin{eqnarray}
{\cal G} (k, i\omega_n) &=& \int_{0}^{\infty}
   \frac{d\omega^{\prime}}{2\pi}
   \left[
   \frac{A(k, \omega^{\prime})}{i\omega_n - \omega^{\prime}}
   + \frac{A(k, -\omega^{\prime})}{i\omega_n + \omega^{\prime}}
   \right]
\end{eqnarray}
we find
\begin{eqnarray}
B(q, \omega) &=& -2\pi \sum_k
   \int_{-\infty}^{\infty} \frac{d\omega^{\prime}}{2\pi}
      A(k+q, \omega^{\prime} + \omega)
      A(k, \omega^{\prime})\nonumber\\
      &&\qquad \times
       [ n_f(\omega^{\prime}) - n_f(\omega^{\prime} + \omega) ]
\end{eqnarray}

We Taylor expand
\begin{eqnarray}
&&\sum_{kq}
   \left[ n_f (\epsilon_{k+q}) - n_f(\epsilon_k) \right]
   \delta(\omega - \epsilon_{k+q} + \epsilon_k) \nonumber\\
   && = - \omega \sum_{kq}
   \frac{\partial n_f}{\partial \omega}
   \delta(\omega - \epsilon_{k+q} + \epsilon_k)  + \ldots
\end{eqnarray}
At $T=0$, $\partial n_f/\partial \omega = - \delta(\omega)$;
and we can eliminate both integrals
\begin{eqnarray}
B(q, \omega) &=& \omega \sum_{k} A(k+q, \omega) A(k,0)
\end{eqnarray}

We now make the assumption that $V_q$ has no significant $q$
dependence and define the function
\begin{eqnarray}
\tilde{B}(\omega) &=& \sum_q \frac{B(q, \omega)}{\omega}
\end{eqnarray}

\noindent
In general $\tilde{B}(\omega)$ will also be a function of $l$,
where $l$ is an index over scattering channels (in 1D the limiting
values $q\approx 0$, $q\approx 2k_f$).  For simplicity we suppress
this
dependence in
what follows.

In terms of $\tilde{B}$ we may write
\begin{eqnarray}
F_2(t) &\approx& - \frac{1}{L} \mid V_0 \mid^2
   \int_0^{\infty} d\tilde{\omega} \tilde{B}(\tilde{\omega})
   \frac{1 - e^{-i\tilde{\omega}t}}{\tilde{\omega}}
\end{eqnarray}

In the physically relevant limit
($\omega \to 0, t\to \infty$), and making the 
substitution $k' = k+q$ we find
\begin{eqnarray}
\tilde{B}(\omega \to 0) &=& \sum_{kk'}
   A(k', \omega \to 0) A(k,0)
   = \left[ \sum_{k} A(k,0) \right]^2
   = n_0^2
\end{eqnarray}
where $n_0$ is the {\it exact} density of states at the Fermi energy.

We impose a soft bandwidth cutoff $\epsilon_0$ (equivalent to having made the 
substitution $\mid V_q \mid^2 \to \mid V_0 \mid^2 e^{-q/q_0} \to
\mid V_0 \mid^2 e^{-\tilde{\omega}/\epsilon_0}$ in \ref{eqn:begin}), 
in which case 
the integral over $\tilde{\omega}$ can be performed exactly to give
\begin{eqnarray}
F_2(t) &=& - \mid V_0 \mid^2 n_0^2 \ln (1 + i\epsilon_0 t) 
\end{eqnarray}

Therefore, 
\begin{eqnarray}
G_h(t) &\approx& -i\Theta(t) e^{-i\omega_T t}
   \frac{1}{(i\epsilon_0 t)^{\alpha_E}}
\end{eqnarray}
where for a $1D$ electron system within these approximations and 
restoring the $2k_f$ scattering channel, the exponent 
$\alpha_E$ is given by
\be
\alpha_E &=& n_0^2 (\mid V_0 \mid^2 + \mid V_{2k_f} \mid^2 ) 
\en
In the special case of free electrons we see that we have recovered 
the Born Approximation to the exact result.

More generally
\be 
\alpha_E &=& \lim_{\omega \to 0} \frac{1}{L}
  \sum_q \mid V_q \mid^2 
  \frac{
  \Im \{ \chi_{\rho}(q, \omega) \}
  }{\omega}
\en
provided that a) the unperturbed Hamiltonian is quadratic in Fermi 
fields b) the leading small $\omega$ behaviour of $\sum_q \mid V_q \mid^2 
\Im \{ \chi_{\rho}(q, \omega) \}$ is linear.

This form of $G_h(t)$ leads directly to the power law assymetries in
core hole lineshapes observed in experiment.   In the case of non--interacting
electrons it can be shown that second order perturbation theory
represents the Born approximation to the the result obtained from
resuming all orders of perturbation theory.
\begin{eqnarray}
\alpha_E &=& \frac{1}{4} \sum_l
  \left[\frac{\delta_l(E_f)}{\pi} \right]^2
\end{eqnarray}

\noindent
where $\delta_l(E_f)$ is the exact phase shift at the Fermi surface in
the $l$'th scattering channel.

It is also possible to calculate the second order shift in the XPS line due to
its interaction with the conduction electrons.   This may be 
evaluated as a conventional self energy correction; the result is 
\be
\Delta E &=& \frac{1}{L} \sum_q \mid V_q \mid^2 
   \Re \{ \chi_{\rho} (q, 0)\}
\en
Gross changes in the susceptibilty due (for example) to the opening 
of a gap in a previously ungapped system, can therefore
lead to a shift in core lines as well as a change in their assymetry.

\section{Core Level coupled to a Luttinger Liquid\label{XPSinLL}}

The following model for a localized single Fermion $d$ (chosen here
to be a
hole)
interacting with a set of Bosons $a_q$
\begin{eqnarray}
\label{eqn:bosf}
H &=& dd^{\dagger} \left[\epsilon_c
  + \sum_q M_q (a_q + a^{\dagger}_{-q}) \right]
  + \sum_q \omega_q a^{\dagger}_q a_q\\
\left\{ d^{\dagger}, d \right\} &=& 1 \qquad
\left[ a_q, a^{\dagger}_{q'} \right] = \delta_{qq'}
\end{eqnarray}

\noindent
may be solved exactly by the canonical transformation
\begin{eqnarray}
\tilde{H} &=& e^s H e^{-s} = \tilde{d}\tilde{d}^{\dagger}
   \left[\epsilon_c - \Delta \right]
   + \sum_q \omega_q \tilde{a}^{\dagger}_q \tilde{a}_q
\end{eqnarray}

\noindent
where
\begin{eqnarray}
s = dd^{\dagger} \sum_q \frac{M_q}{\omega_q}
   (a^{\dagger}_q - a_q) \qquad
\Delta &=& \sum_q \frac{M_q^2}{\omega}
\end{eqnarray}

\noindent
This model has the unphysical feature that the interaction term
does not conserve momentum -- the core level does not recoil when
electrons are scattered from it.

It is also possible to find the fermion correlation function
\begin{eqnarray}
iG_d(t) &=& \langle T d^{\dagger}(t) d \rangle
\end{eqnarray}

\noindent
in closed form.  We consider only the case $t>0$, for which
the result is
\begin{eqnarray}
\label{eqn:Gc}
iG_d(t) &=& \left[1 - n_f(\epsilon_c - \Delta)\right]
   \exp \left[ -\phi(t) \right]\\
\phi(t) &=& \sum_q \left( \frac{M_q}{\omega_q} \right)^2
   \left[N_q\left(1 - e^{-i\omega_q t}\right)
   + \left(N_q + 1\right) \left(1 - e^{i\omega_q t}\right)
   \right]
\end{eqnarray}

\noindent
where $N_q$ is a boson occupation factor, which we set to zero in what
follows.  The procedure for solving the model and obtaining
correlation functions is described in some detail in
\cite{mahan-schotte}.

The simplest prototypical Hamiltonian for a core level coupled to
interacting conduction electrons in 1D may be written
\begin{eqnarray}
H_1 &=& \sum_k \epsilon_p c^{\dagger}_p c_p
   + \frac{1}{2L} \sum_q V_q \rho(q) \rho(-q)\\
V_{IMP} &=& \frac{1}{L} \sum_q M_q \rho(q) dd^{\dagger}\\
\rho(q) &=& \sum_p c^{\dagger}_{p-q}c_q \qquad
\left\{d, d^{\dagger}\right\} = 1
\end{eqnarray}

\noindent
where $\epsilon_p$ is the free electron dispersion and
$V_q$ the FT of the electron--electron interaction.  This model
(the Tomonaga Model) may be solved by linearizing the
electron dispersion $\epsilon_p = v_f(|p| - p_f)$
and introducing a set of collective Bosonic
coordinates to describe excitations of the conduction electrons.

We now make the observation that the bosonized form of this model
is (up to a constant) in {\it exact} correspondence with model of a
single fermion interacting with a set of bosons solved above.

Explicitly, we split the density operator into separate parts
referring to right and left moving electrons
$\rho (q) = \rho_1(q) + \rho_2 (q)$
which we normalize as bosonic operators
\begin{eqnarray}
b_q &=& \frac{\pi}{L\pi}
\left[\theta(q) \rho_1(q) + \theta(-q) \rho_2(q)\right]\\
\left[ b_q, b^{\dagger}_{q'} \right] &=& \delta_{qq'}
\end{eqnarray}

\noindent
We write the kinetic energy term as
\begin{eqnarray}
\sum_{q}\omega_q b^{\dagger}_q b_q \qquad \omega_q = v_f \mid q \mid
\end{eqnarray}

\noindent
and make use of the canonical transformation
\begin{eqnarray}
b_q + b^{\dagger}_{-q} &=&
   \sqrt{\frac{\omega_q}{E_q}}(\alpha_q + \alpha^{\dagger}_{-q})
   \qquad
   \left[ \alpha_q, \alpha^{\dagger}_{q'} \right] = \delta_{qq'}\\
E_q &=& \sqrt{\omega_q^2 + 4\omega_q \overline{V}_q}
   = v_f \mid q\mid \sqrt{1 + \frac{2V_q}{\pi v_f}}
\end{eqnarray}

\noindent
to obtain
\begin{eqnarray}
H &=& \sum_q E_q \left(\alpha^{\dagger}_q \alpha_q
   + \frac{1}{2} \right)
+ dd^{\dagger} \left[ \sum_q \overline{M}_q
   (\alpha_q + \alpha^{\dagger}_{-q}) \right]
   + \epsilon_d dd^{\dagger}
\end{eqnarray}

\noindent
which is clearly of the same form as \eref{eqn:bosf}.

The density of states at the Fermi energy in a Luttinger Liquid is
identically zero and one might expect, on the basis of the arguments
presented in Appendix D that this would mean that it would present no
asymmetry in core hole responses.   This expectation is not
born out --- infact there is {\it no} essential difference between the
bosonized
Hamiltonian for free electrons coupled to a core level and that for
interacting electrons coupled to a core level (provided that
we neglect all ``backscattering'' processes).  The response of the
interacting system is therefore qualitatively the same.   However,
since the Bosonic eigenstates of the interacting electron system are
not
exactly the same as those for the free one, we must remember to modify
the energy levels and matrix elements in the expression for $G_d(t)$
\ref{eqn:Gc}.
\begin{eqnarray}
\omega_q &\to& E_q\\
M_q &\to& \sqrt{\frac{\omega_q}{E_q}} \frac{M_q \mid q \mid}{2 \pi}
\end{eqnarray}

The all important logarithm in the response of the electron system to
the
impurity may be obtained by replacing the remaining sum over $q$ with
an integral.  In the case where all interactions are
$\delta$ functions it is particularly easy to understand what happens.

Then,
\begin{eqnarray}
\phi(t) &=& \int_0^{q_0} \frac{dq}{2\pi}
   \left( \frac{\overline{M}_q}{E_q} \right)^2
   \left[1 - e^{-iE_q t}\right]\\
E_q &=& \overline{v}\mid q\mid \qquad \overline{M}_q =
\frac{M|q|}{2\pi}
\end{eqnarray}

\noindent
where $\overline{v} = v_f\sqrt{1 + 2V/\pi v_f}$ and $q_0$ is a band
cutoff.  We may bring this into the from of the non--interacting case
by the
change of variables $\overline{q} = (v_f/\overline{v}) q$.  Then,
without
needing to evaluate the integral, we know that the asymmetry exponent
will be
modified according to
\begin{eqnarray}
\alpha_E (V) &=& \left(\frac{v_f}{\overline{v}}\right)^3
   \alpha_E (0)\\
\left(\frac{v_f}{\overline{v}} \right)^3 &=&
   \left(1 + \frac{2V}{\pi v_f}  \right)^{-\frac{3}{2}}
\end{eqnarray}

\noindent
i.e. suppression for $V>0$ and enhancement for $V<0$.

A more careful analysis of the $q \to 0$ limit of integrand in the
general
case leads to the result \eref{eqn:LLalpha} in terms of the Luttinger
Liquid
correlation parameter of the quoted in the text.   The problem
of the X--ray response of a LL has been considered previously by 
a number of authors \cite{alpha}.

It is clear that the approach of Appendix 1 fails dramatically
and it is worth pausing to consider why.  The result
\eref{eqn:theresult}
is quite generally valid at this order.  The error arises in
the evaluating density--density correlation function $\Pi(q,
i\Omega)$
as a single loop of fermions.  In the case
of a system with electron--like quasiparticles this is unlikely to
lead to a large qualitative error, but in the case of a Luttinger
liquid where Wick's Theorem does not hold it is quite simply the wrong
thing to do.  Fortunately it is easy to evaluate the {\it exact} form
of
$\Pi(q, i\Omega)$ by bosonization.  Substituting this in
\eref{eqn:theresult}
leads to the same results as the approach outlined above.

We note that none of these arguments are affected by the
introduction of spin, and may even be applied to an ``insulating''
phases, if the gap is in the spin rather than the charge sector
(Luther--Emery liquid).

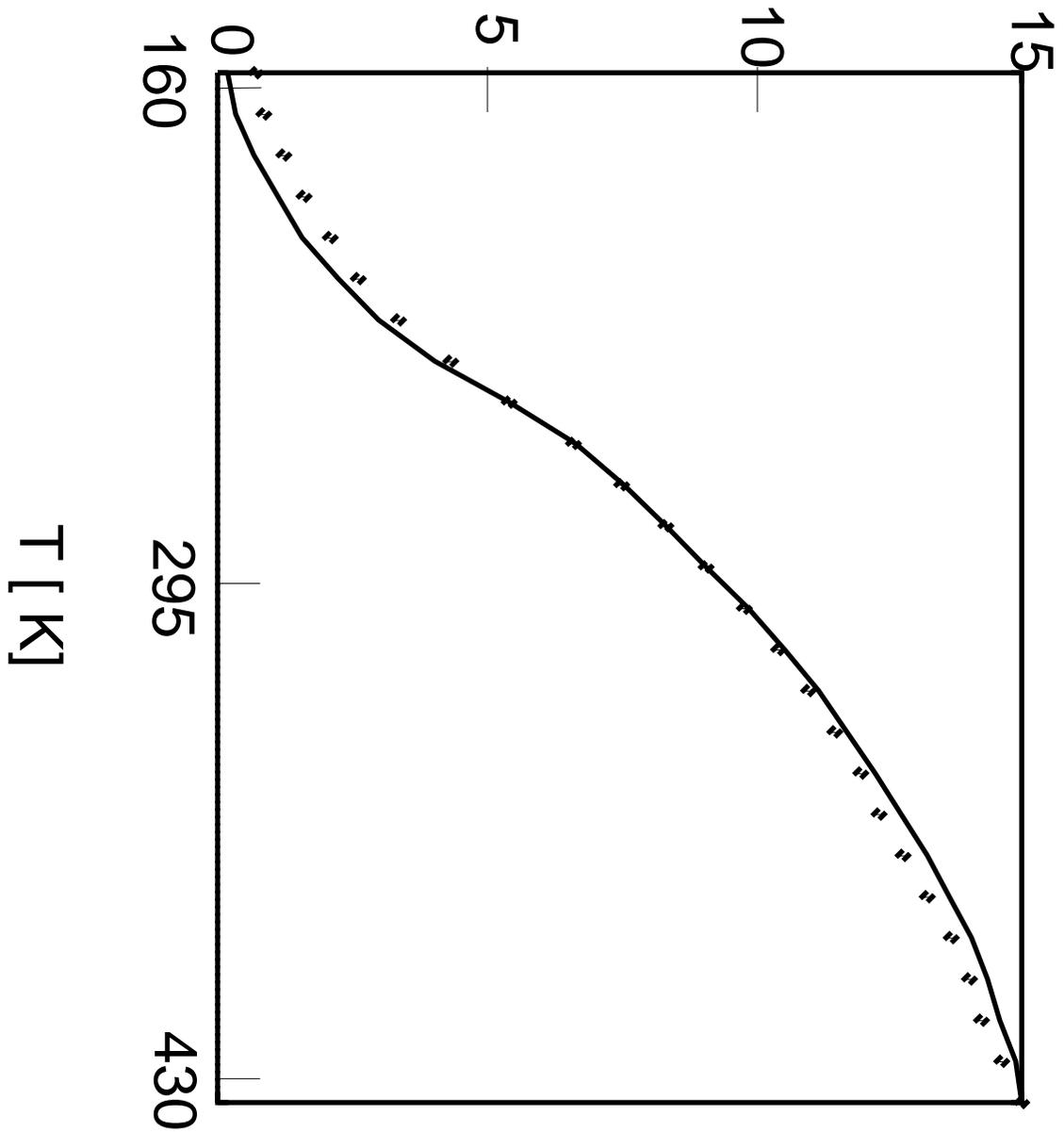
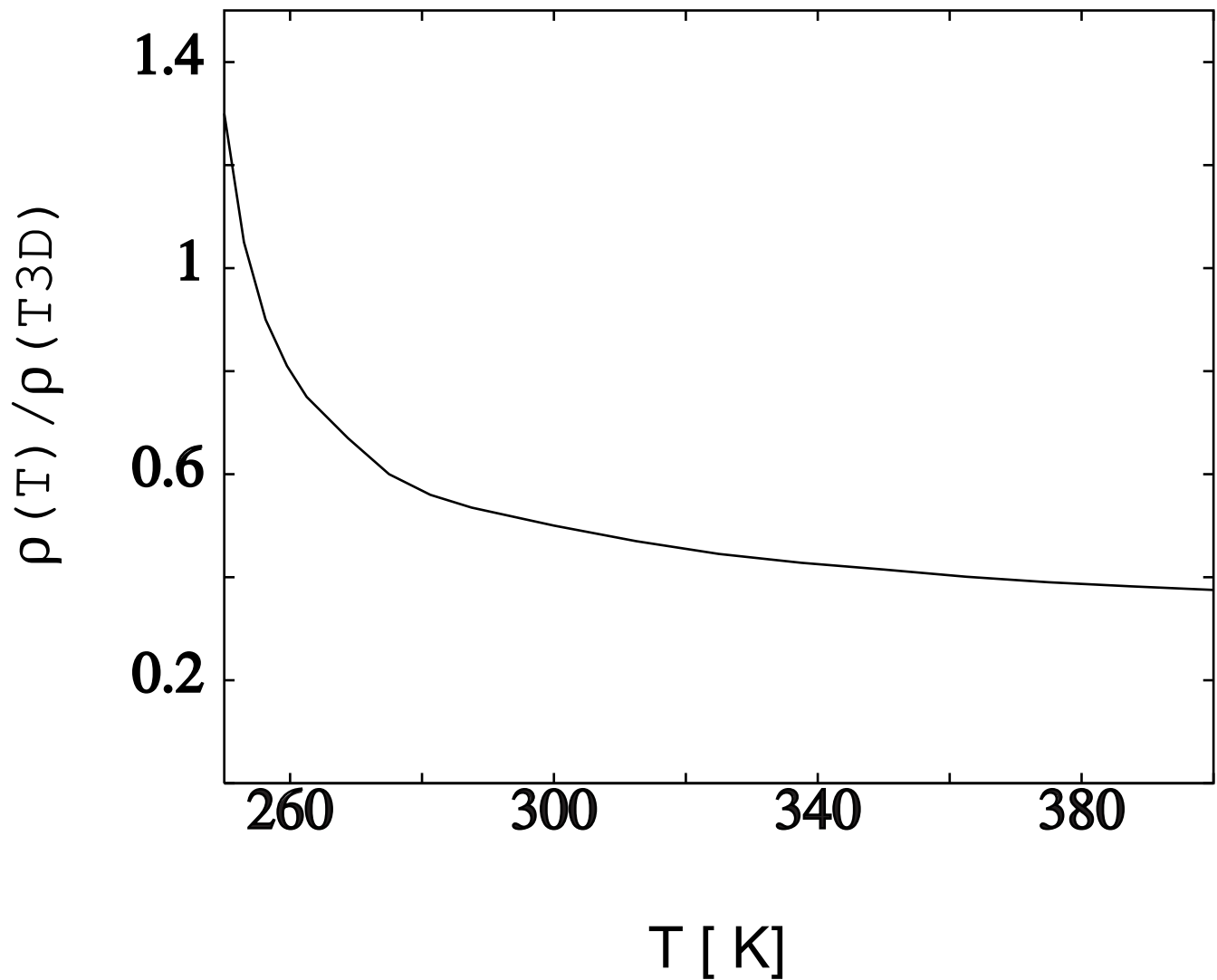
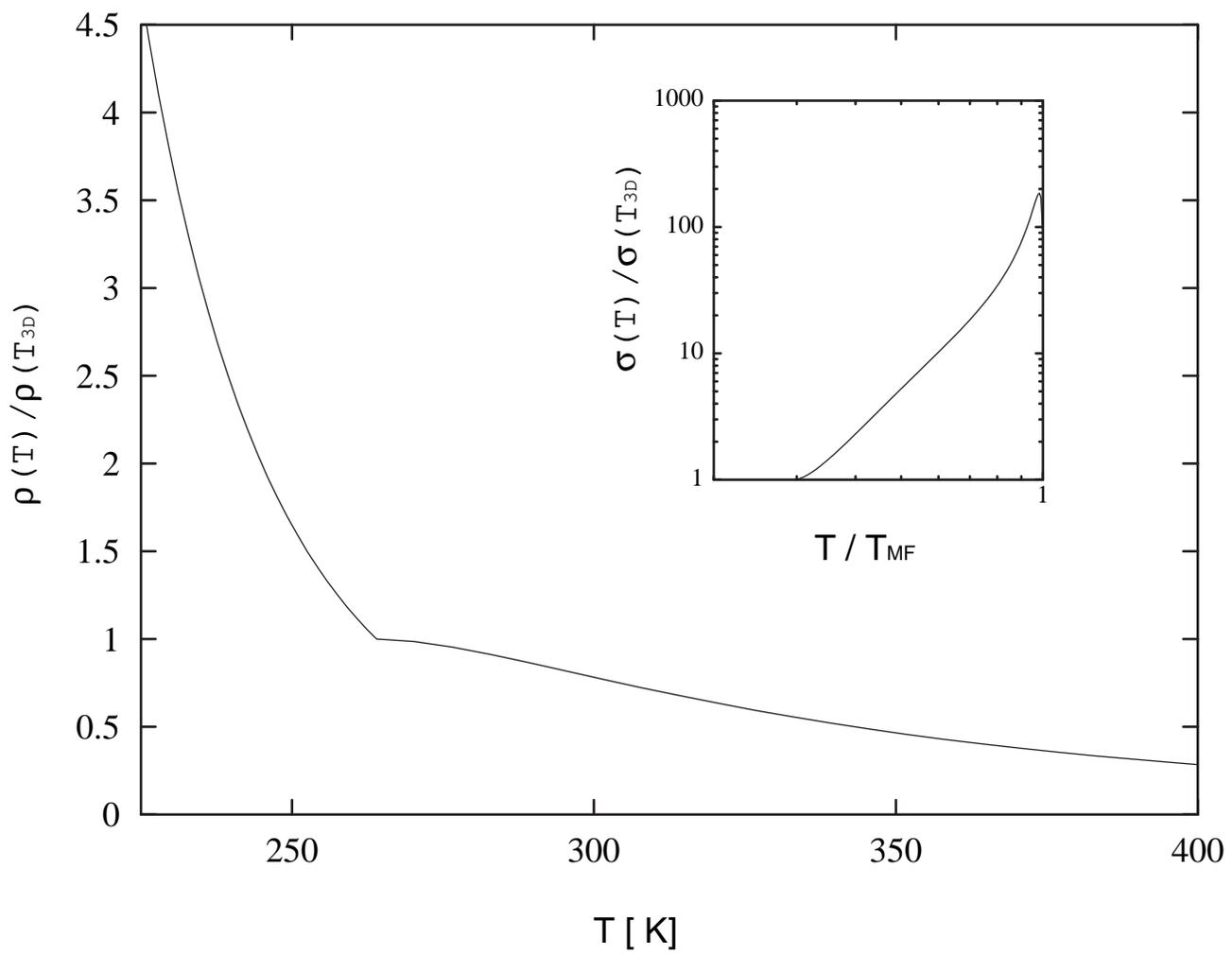
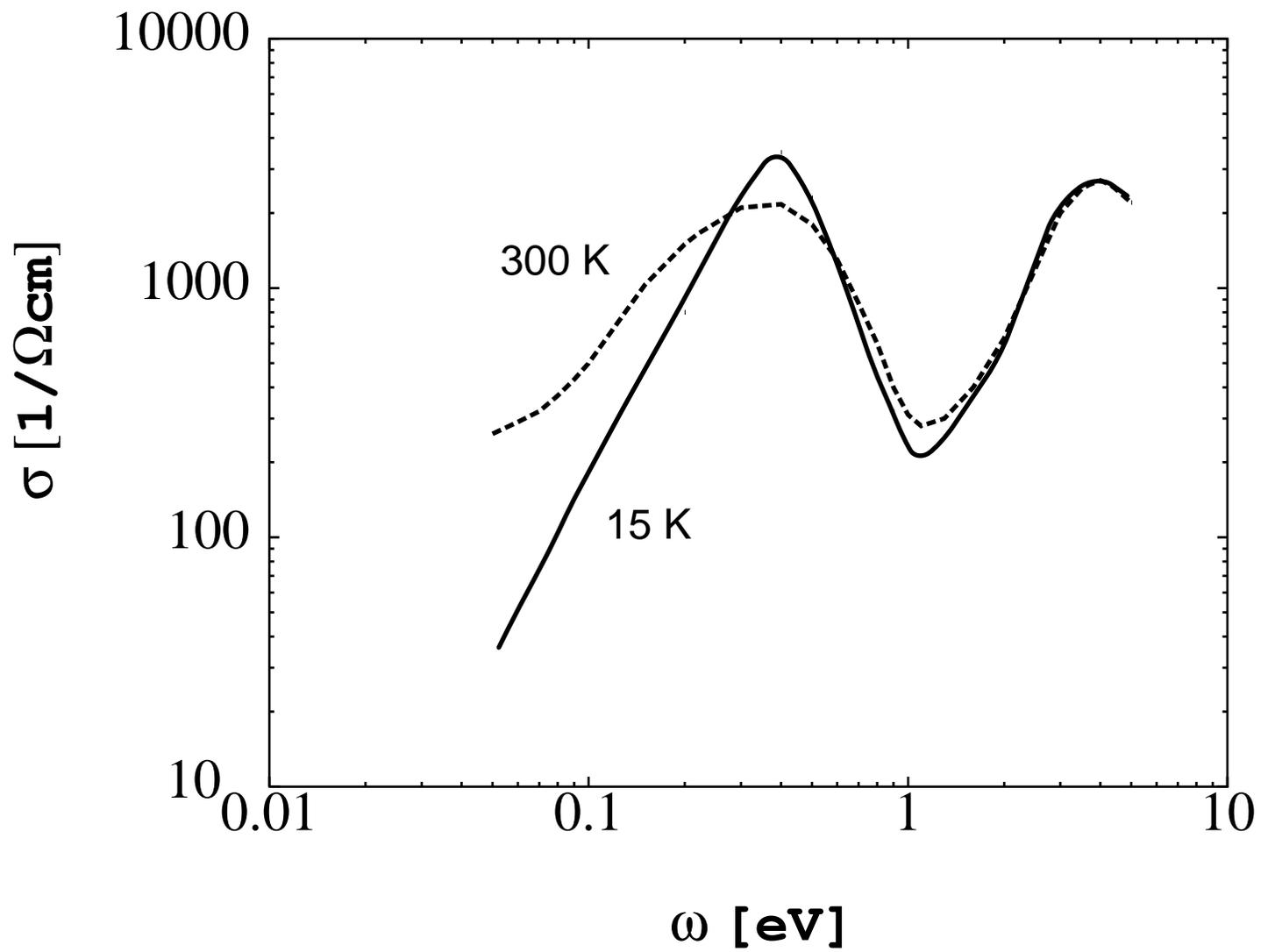
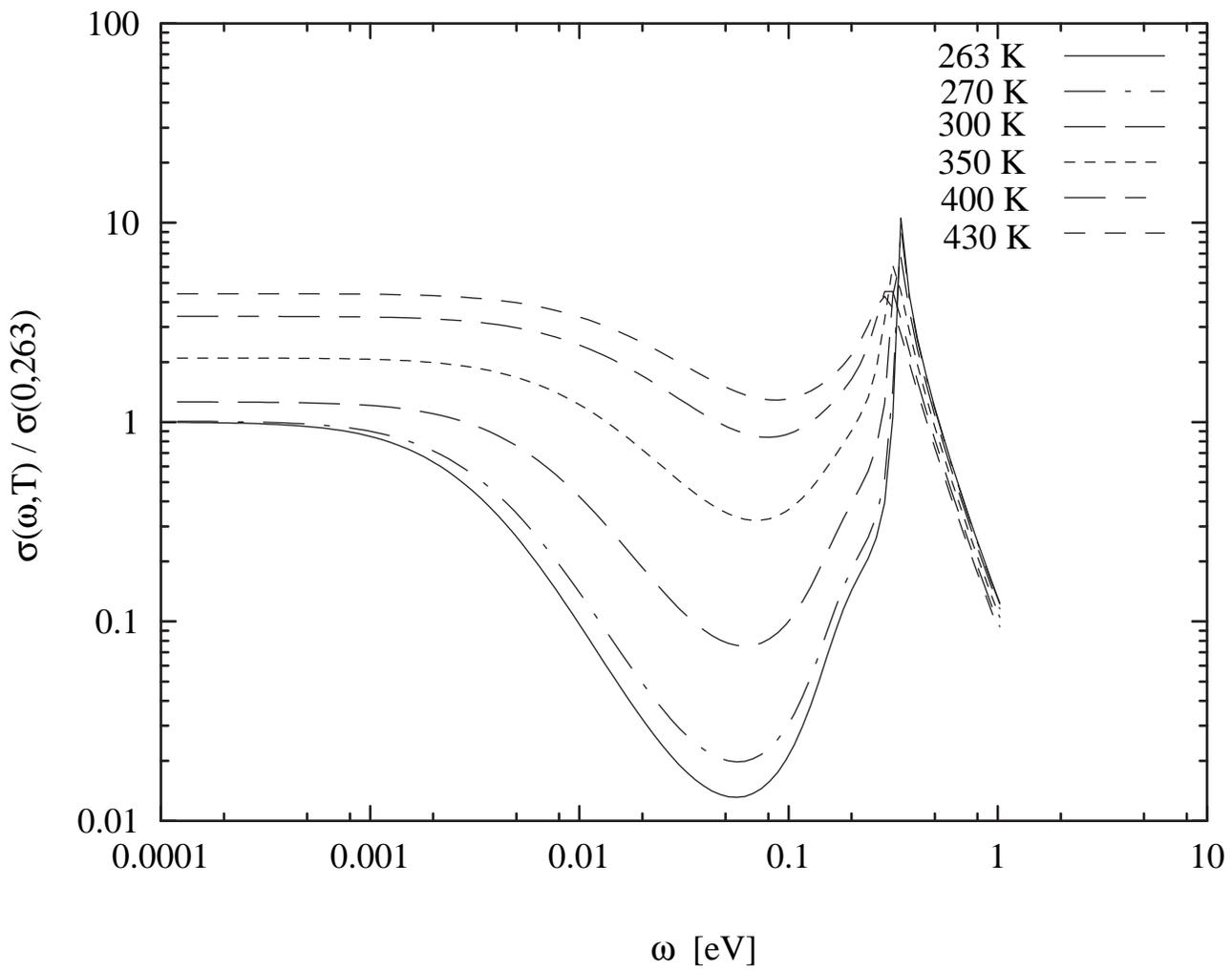
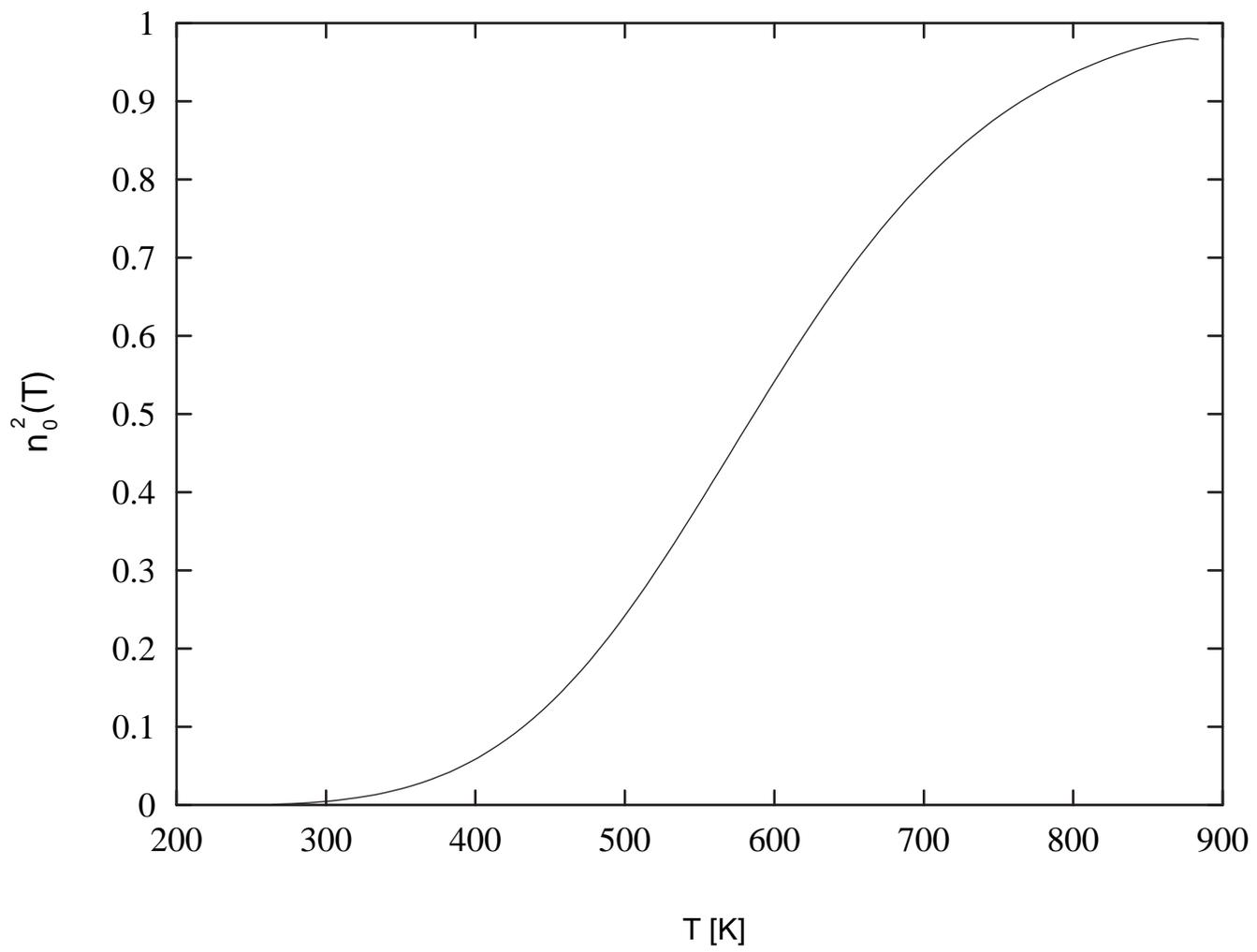
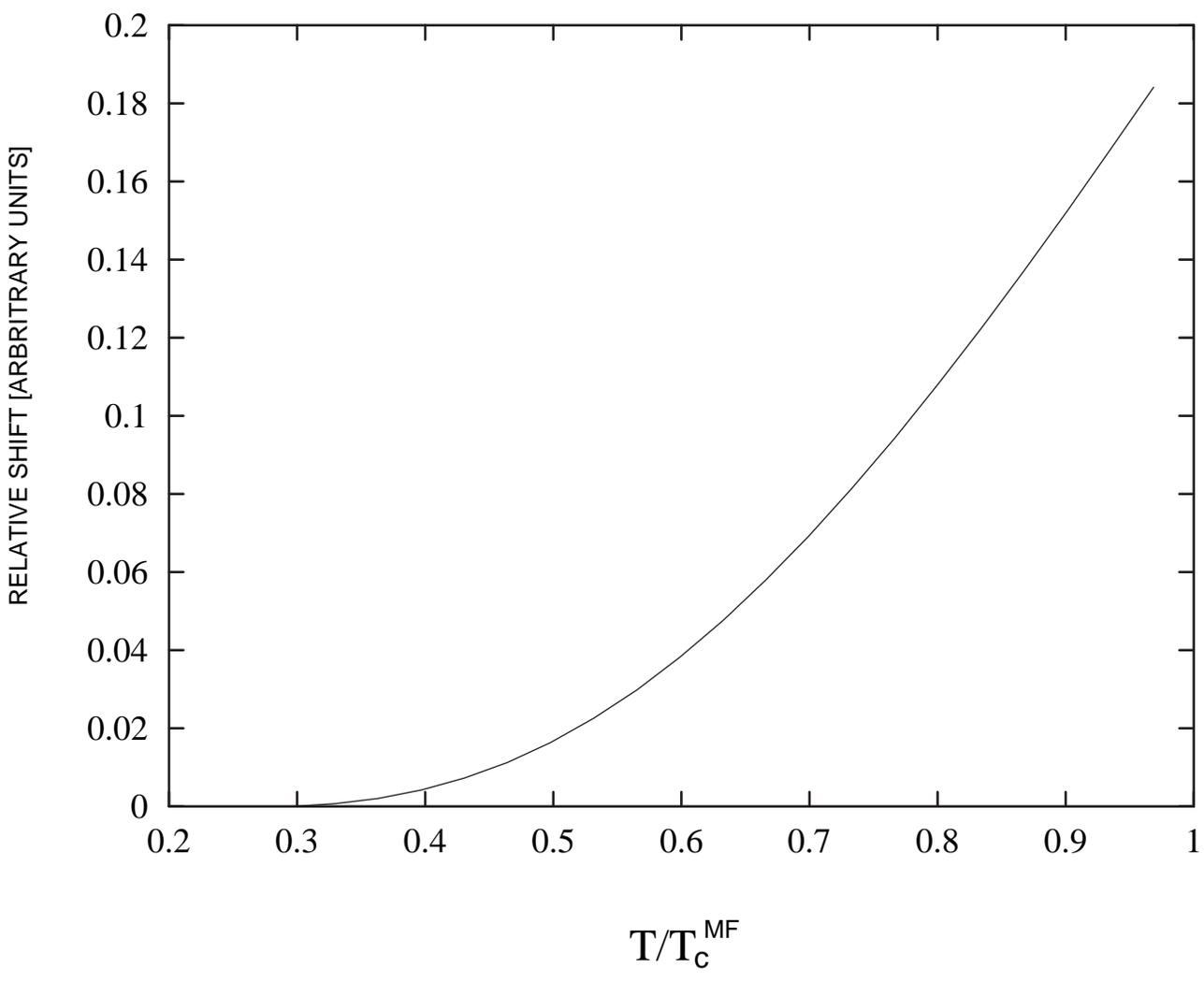
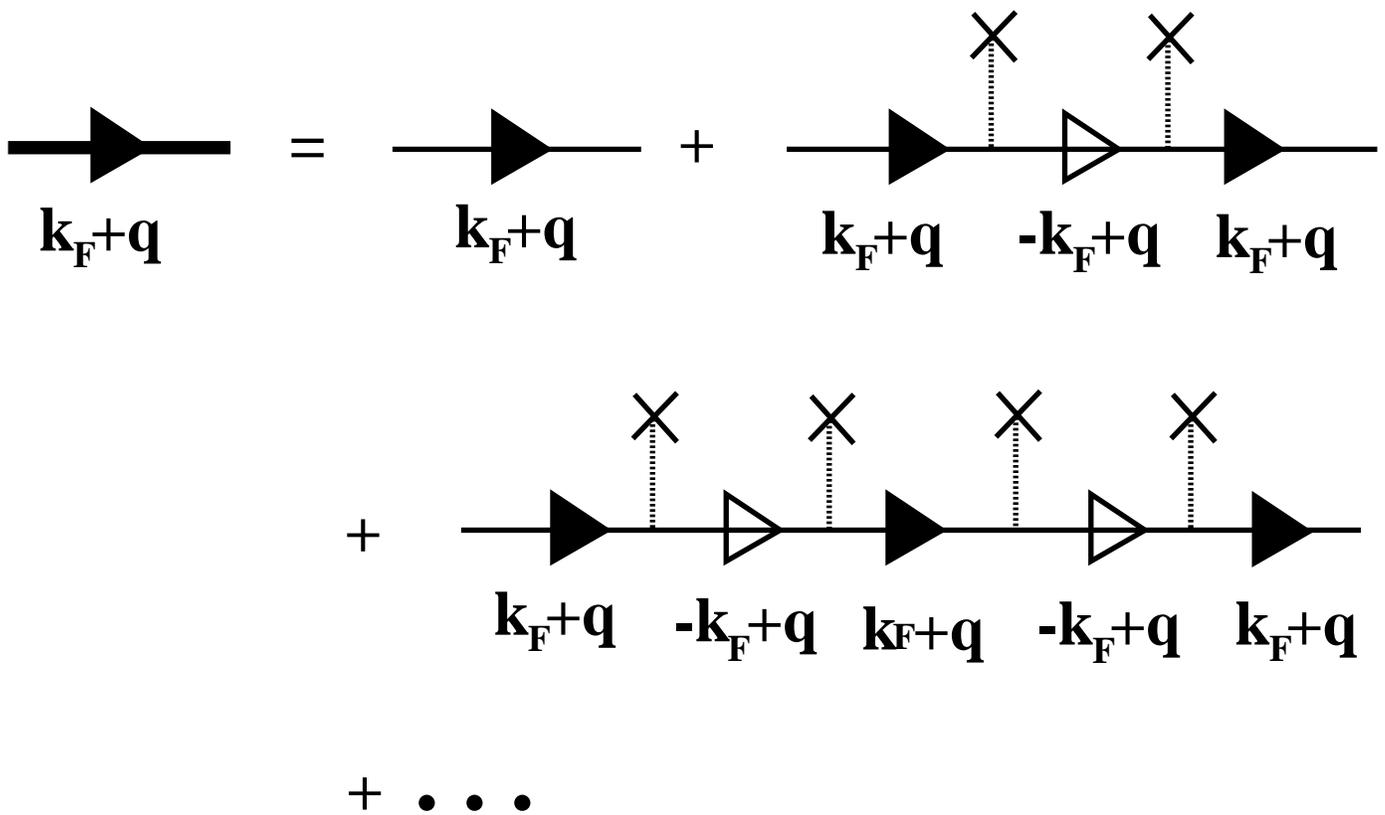
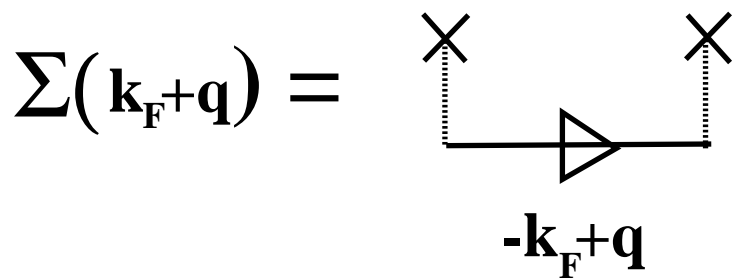
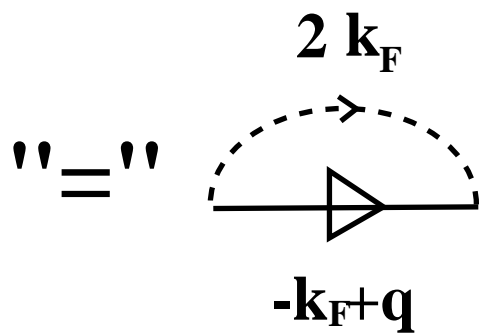
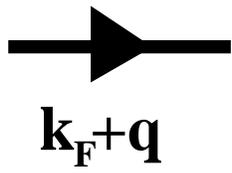
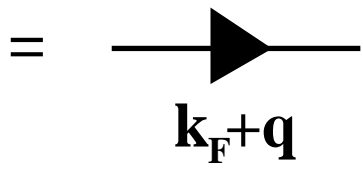
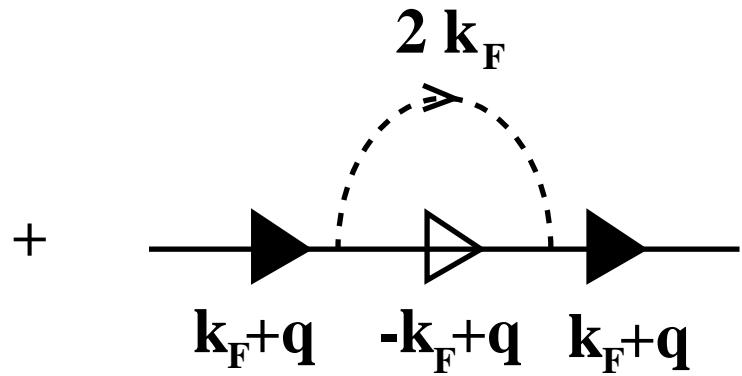
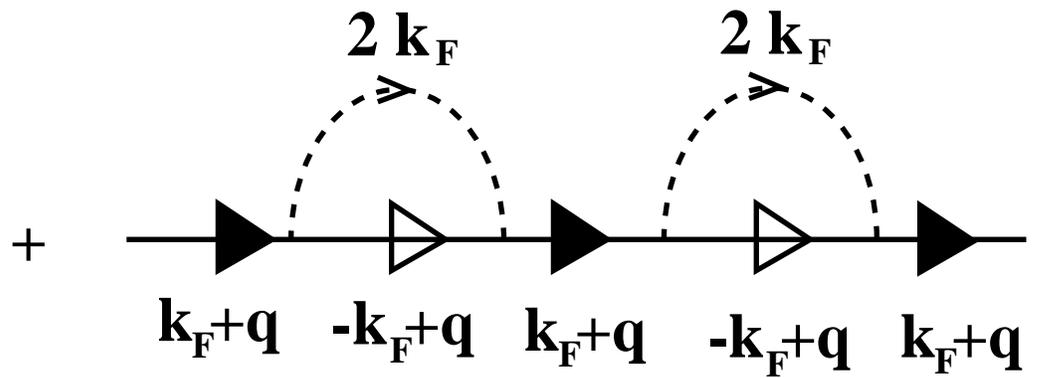
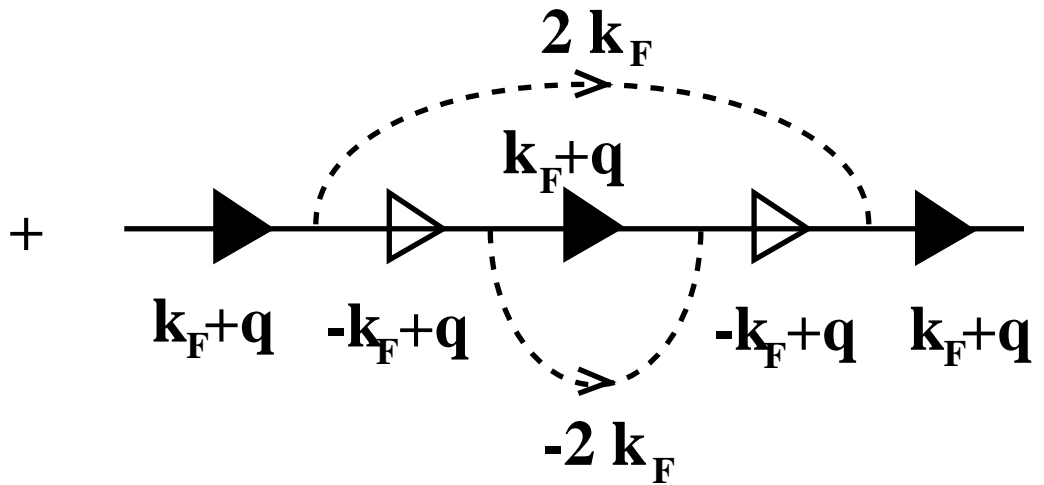
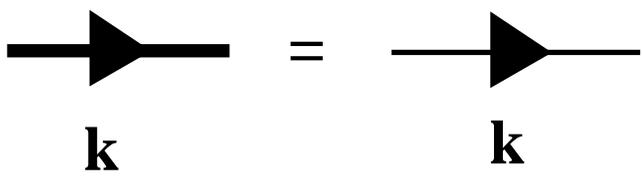
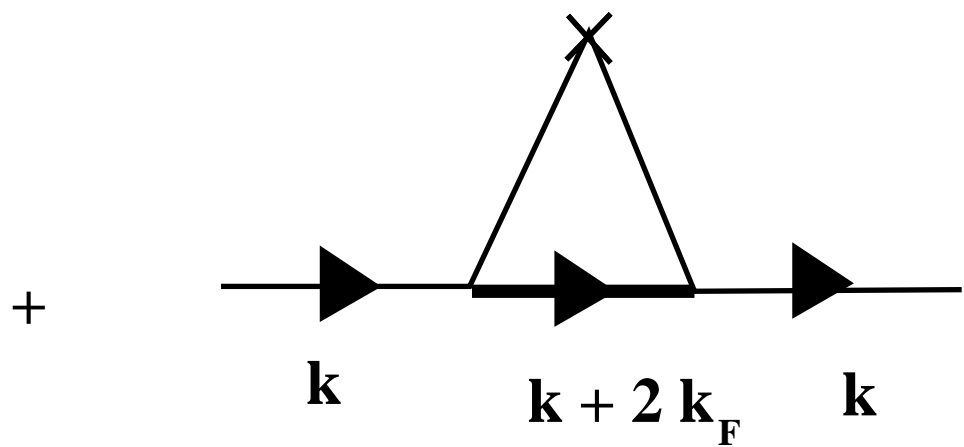
\end{document}